\documentclass[journal]{new-aiaa}
\usepackage[utf8]{inputenc}
\usepackage{textcomp}

\usepackage{graphicx}
\usepackage{amsmath}
\usepackage[version=4]{mhchem}
\usepackage{siunitx}
\usepackage{longtable,tabularx}
\usepackage{hyperref}
\hypersetup{
    hidelinks
}

\usepackage{url}
\usepackage{subcaption}
\usepackage{enumitem}
\usepackage{amsmath}
\usepackage{bm}

\usepackage{multirow}
\usepackage{ulem}

\usepackage[title]{appendix}
\usepackage{array}   

\usepackage{lipsum} 
\usepackage{xcolor} 
\usepackage{todonotes}

\usepackage{lineno} 

\usepackage[version=4]{mhchem}
\usepackage{siunitx}
\usepackage{longtable,tabularx}
\title{
Optimizing flow control with ensemble Kalman method for mitigating flow-induced vibration
}

\author{Yi Liu \footnote{Postdoctoral Research Associate, LNM, Institute of Mechanics; liuyi@imech.ac.cn }}

\author{Shizhao Wang \footnote{Professor, LNM, Institute of Mechanics; wangsz@lnm.imech.ac.cn }}

\author{Xin-Lei Zhang \footnote{Associate professor, LNM, Institute of Mechanics; zhangxinlei@imech.ac.cn (Corresponding author).}}

\author{Guowei He \footnote{Professor, LNM, Institute of Mechanics; hgw@lnm.imech.ac.cn.}}

\affil{State Key Laboratory of Nonlinear Mechanics, Chinese Academy of Sciences, 100190 Beijing, \\ People’s Republic of China}

\affil{University of Chinese Academy of Sciences, 100049 Beijing, People’s Republic of China}

\begin{document}

\maketitle

\begin{abstract}
    The ensemble Kalman method is introduced for optimizing flow control strategies in order to mitigate the flow-induced vibration of structures.
    Different types of control strategies such as passive control, open-loop active control, and closed-loop active control are tested, showing the flexibility of the method in flow control optimization.
    The ensemble Kalman method is first tested to mitigate vortex shedding of flows around a circular cylinder by optimizing the placement of small cylinders downstream.
    Further, the method is assessed to suppress shock buffeting over the NACA 0012 airfoil by optimizing the movement of a compliant aileron.
    Our results for all test cases show that the ensemble-based method can effectively find optimal control strategies that significantly reduce the vibrations of aerodynamic force,
    and can be a useful alternative for flow control optimization, due to its merits in non-intrusiveness and ease of implementation. 
\end{abstract}

\section*{Nomenclature}

{
\renewcommand\arraystretch{1.0}
\noindent\begin{longtable*}{@{}l @{\quad=\quad} l@{}}
$C_d$   & drag coefficient \\
$C_f$   & friction coefficient \\
$C_l$   & lift coefficient \\
$C_p$   & pressure coefficient \\
$C_{P_b}$ & mean base pressure coefficient \\
$C$     & chord length \\
$d, D$   & diameter of the small and main cylinder \\
$E$     &  total energy   \\
$F_d,F_l$     &   drag and lift force  \\
$H$          &  total enthalpy  \\
$L_\text{ref}$             &  reference length \\
$M$             &  number of samples \\
$Ma$            &  Mach number    \\
$N$             &  number of mesh cells \\
$p$             &  order of convergence in GCI method \\
$\mathsf{P}$    &  model error covariance   \\
$Pr$            &  Prandtl number  \\
${q_j},q_j^{(\text{t})}$  &  laminar heat flux and turbulent heat flux  \\
$r_\text{eff}$  &  effective mesh refinement ratio  \\
$\mathsf{R}$    &  observation error covariance  \\
$Re$  & Reynolds number \\   
${\mathbf{S}},{\Omega}$  & strain-rate tensor and rotation-rate tensor \\  
$S_\text{flap}$  &  area of swept by the camber of aileron during one cycle of flapping \\
$St$   &  Strouhal number  \\ 
$t$    &  time  \\
$T$    &  static temperature  \\
${T_\infty }$ & temperature at far-field \\  
${U_\infty }$   &  free-stream velocity  \\ 
${\mathbf{u}}$  &  velocity vector \\
$w$       &  weights of neural networks \\
$W$       &  ensemble of weights, $W=\{ w \}_{m=1}^M$ \\
$x$       &  Cartesian coordinates \\
$y^+$     &  nondimensional wall distance \\
$\alpha$  &  angle of attack  \\
$\delta_{ij}$  & Kronecker delta function  \\
$\Delta$       & Displacement of the trailing edge \\
$\mathcal{H}$     &  model operator that maps the neural network weights to the observed quantities \\
${\mu, \mu _t }$  &  dynamic viscosity and turbulent viscosity  \\
$\epsilon_\text{rel}$  & relative change \\
${\theta}$     &  circumferential coordinate \\
${\eta}$     & ratio of aileron oscillation frequency to shock wave frequency \\
$\rho$            &  fluid density  \\
$\sigma$     &  viscous stress \\
$\tau$            &  Reynolds stress  \\
$\omega$      &  frequency \\
$\boldsymbol{\omega}$   &   vorticity  \\
${\nabla}$  &  gradient operator \\
$\|*\|$       &  L2 norm  \\

\multicolumn{2}{@{}l}{Superscripts}\\
${(i)}$ & index of tensor basis \\
$n$ & index of iteration \\
${\top}$ & transpose \\

\multicolumn{2}{@{}l}{Subscripts}\\
${i,j,k}$  & 1, 2, 3, tensor indices \\ 
${m}$ & index of sample \\
$\text{mean}$ & mean \\
$\text{obj}$  & objective  \\
$\text{rms}$  & root-mean-square \\
$\text{sep}$  & flow separation \\
$\text{std}$  & standard deviation \\
${\infty}$    & at infinity \\

\end{longtable*}}

\section{Introduction}
Flow control is of practical interest in extensive applications spanning from aerospace engineering to environmental science. 
It can significantly improve device performance, such as drag reduction~\cite{endrikat2021influence,haffner2022drag}, shock buffeting suppression~\cite{giannelis2017review}, and mixing enhancement~\cite{depuru2015vortex}, by manipulating flow patterns.
Particularly, 
flow control techniques can be used to reduce flow-induced vibrations, which can prolong the fatigue life of structures and prevent catastrophic aeroelastic failures~\cite{cattafesta2008active,AFResearch2020,zhu_large-eddy_2022}.
To this end, it is worth developing effective flow control strategies for mitigating flow-induced vibrations.

Over the past few decades, various research has focused on developing effective flow control strategies~\cite{stanewsky2001adaptive,bewley2001flow,gardner2023review,guan2019open}, 
which are categorized into passive and active control strategies, depending on whether external input energy is required~\cite{gad2003flow}.
Passive flow control techniques, such as riblets and vortex generators, manipulate flow patterns by modifying geometric surfaces, which do not require external energy. 
For instance, riblets with optimized groove size and shape can reduce skin friction drag by altering near-wall turbulence~\cite{endrikat2021influence}.
Similarly, vortex generators can reduce the size of separation vortices by alleviating shear layer instability~\cite{pickles2020control}.
In contrast, active flow control techniques involve the use of external energy to manipulate the flow with actuators, e.g., trailing-edge flapping~\cite{feszty2004alleviation}.
This method can generate trailing-edge vortices (TEV), which interact with the dynamic-stall vortex to effectively reduce negative pitching moments and aerodynamic damping.
The active flow control can be implemented in either an open-loop or closed-loop manner.
Open-loop control utilizes predetermined parameters, e.g., the flapping frequency, to achieve desired flow responses, while closed-loop control dynamically adjusts actuation based on real-time feedback from sensors~\cite{collis2004issues}.
The effectiveness of these control strategies highly depends on the corresponding control parameters such as the frequency and magnitude of the trailing edge flapping~\cite{monastero2019effect}.
Hence, finding optimal control parameters becomes critical for each flow control method.

Designing effective flow control strategies is often a challenging endeavor as it involves exploring a large parameter space~\cite{king2005nonlinear, cho2011adaptive, brunton2020machine, gardner2023review, okbaz2023flow}.
A typical case is flow control with the vortex generator, which requires optimizing design parameters, e.g., installation positions~\cite{chanzy2020optimization,pickles2020control}, spatial distribution~\cite{siconolfi2015stability}, and shape~\cite{barter1995reduction,jirasek2006design} of the generator. 
Also, flow control with trailing-edge flapping needs to optimize aileron length~\cite{wan2012hovering} and the amplitude~\cite{krzysiak2006aerodynamic}, frequency~\cite{doerffer2010unsteady}, and phase difference~\cite{lee2011unsteady} of flap deflection, thereby suppressing dynamic stall and transonic shock buffeting.
The classical method for designing flow control strategies needs to explore the flow dynamics across a wide range of parameters, followed by using physical insights to develop specific control strategies. 
Hence, it often demands considerable effort to achieve effective outcomes.
Furthermore, the nonlinear effects become pronounced for flows at high Reynolds numbers, exemplified by typical limit-cycle behaviors such as transonic shock buffeting~\cite{lee1990oscillatory,giannelis2017review}. 
Such nonlinear effects also pose considerable difficulties for classical methods to design effective flow control techniques~\cite{barbagallo2012closed,aleksic2014need}.
Alternatively, the optimization-based flow control method can effectively design flow control strategies by exploring large parameter spaces.
It is achieved by formulating flow control as optimization problems that aim at minimizing or maximizing specific flow properties~\cite{kirk2004optimal}. 
In doing so, one can search for optimal control parameters in high dimensional space.
The optimization-based flow control methods have been used to reduce recirculation in separated boundary layers 
in high-dimensional flow control problems~\cite{passaggia2018optimal,repolho2022active}. 
Furthermore, such methods are used to 
control transonic shock wave/boundary layer interactions (SWBLI), which involve complex nonlinear interactions between jets, vortices, and shock waves~\cite{chanzy2020optimization}. 
These works demonstrate the effectiveness of the optimization-based flow control method for high-dimensional and nonlinear control problems.

The optimization-based flow control aims to minimize a cost function that measures the objective quantities such as the vibration of lift force.
It typically resorts to adjoint methods~\cite{jameson1988aerodynamic,gunzburger2002perspectives} that solve the adjoint equations to provide the gradient of the cost function, guiding the search for optimal control parameters.
Such methods have been widely used in optimization-based flow control to suppress vortex shedding~\cite{hill1992theoretical}, reduce disturbance growth in boundary layers~\cite{pralits2002adjoint}, delay stall for the NACA 4412 airfoil~\cite{nemili2017accurate}, enhance turbulence mixing~\cite{onder2016optimal}, reduce aerodynamic noise~\cite{zhou2021efficient}, and so on.
However, the adjoint method requires significant memory for storage as it needs to store all snapshots of the flow field.
This would also cause substantial computational costs due to the need to iteratively solve a large linear system (i.e. the adjoint equation)~\cite{mura2017efficient,paris2021robust}. 
Additionally, the adjoint method requires extra effort to develop the adjoint solver, often necessitating significant code modifications, particularly for legacy codes~\cite{zhang2020regularized,duraisamy2021perspectives}. 
Therefore, it is of practical interest to develop non-intrusive alternatives for optimization-based flow control.

The ensemble Kalman method~\cite{assimilation2009ensemble, zhang2020regularized} is a non-intrusive optimization method that leverages the sample covariance between inputs and outputs from multiple primal simulations to approximate the gradient of cost functions.
The method is typically used for solving data assimilation problems, i.e., assimilating observation data into dynamic systems to provide the best estimate of the system state.
Iglesia et al.~\cite{iglesias2013ensemble} propose using the ensemble-based method to solve the inverse problems.
Due to its non-intrusiveness, the ensemble method does not require modifications to the CFD solver, making it straightforward to implement.
For this reason, the method is flexible for various objective functions~\cite{zhang2021assimilation} without the need to develop specific algorithms in gradient computation.
Moreover, the capability of the method has been demonstrated in handling high-dimensional and nonlinear inverse problems~\cite{zhang2020evaluation,zhang2024parallel}, allowing it to address flow control problems with large parameter spaces and nonlinear effects.
The ensemble Kalman method has been successfully applied to a wide range of engineering problems, such as acoustic inversion for jet noise prediction~\cite{zhang2022acoustic}, reconstruction of unsteady viscous flows~\cite{mons2016reconstruction}, and turbulence modeling for compressible flow around the M6 wing~\cite{liu2023learning} and incompressible flow over an axisymmetric body of revolution~\cite{liu2024data}.
However, the feasibility of the ensemble method for flow control still lacks investigation.

This work aims to investigate the application of the ensemble Kalman method for flow control in mitigating flow-induced vibrations.
The method is used to optimize control parameters for different flow control strategies, including passive control, open-loop active control, and closed-loop active control.
The capability of the ensemble-based method is tested for flow control optimization in two flow applications.
The first case is low-speed flows past a circular cylinder at $Re_D = 3900$, where the method is applied to suppress vortex shedding by optimizing the placement of small cylinders downstream.
The second case is the transonic buffeting flow over a NACA 0012 airfoil, where the ensemble method is used to mitigate the shock buffeting by optimizing the movement of a compliant aileron in both open-loop and closed-loop manners.
This work highlights the flexibility of the ensemble-based method for optimizing different flow control strategies due to its derivative-free nature.

The rest of the paper is outlined as follows.
The numerical solver and the methodology of the ensemble Kalman method are introduced in Section~\ref{sec:2-NumericalMethod}.
The test cases and corresponding results are presented and analyzed in Section~\ref{sec:3-NumericalResults}.
Finally, this paper is concluded in Section~\ref{sec:4-Conclusion}.

\section{Methodology \label{sec:2-NumericalMethod} }

\subsection{Flow simulation}
\label{sec:2.1-FlowSolver} 
    In this work the flow control is conducted in the simulation environment.
    The unsteady Reynolds averaged Navier-Stoke (RANS) method is used to emulate flow dynamics.
    The unsteady RANS equations for compressible flows can be written as~\cite{blazek2015computational}
    \begin{subequations}
    \label{eq:NSEquations}
    \begin{equation}
        \frac{{\partial \rho }}{{\partial t}} + \frac{{\partial (\rho {u_j})}}{{\partial {x_j}}} = 0 ,\label{subeq:1}
    \end{equation}
    \begin{equation}
        \frac{{\partial (\rho {u_i})}}{{\partial t}} + \frac{\partial }{{\partial {x_j}}}(\rho {u_i}{u_j}) =  - \frac{{\partial p}}{{\partial {x_i}}} + \frac{{\partial {\sigma _{ij}}}}{{\partial {x_j}}} + \frac{{\partial {\tau _{ij}}}}{{\partial {x_j}}},\label{subeq:2}
    \end{equation}
    \begin{equation}
        \frac{{\partial (\rho E)}}{{\partial t}} + \frac{\partial }{{\partial {x_j}}}(\rho H{u_j}) = \frac{{\partial ({\tau _{ij}}{u_i})}}{{\partial {x_j}}} + \frac{{\partial ({\sigma _{ij}}{u_i})}}{{\partial {x_j}}} - \frac{{\partial {q_j}}}{{\partial {x_j}}} - \frac{{\partial q_j^{\left( t \right)}}}{{\partial {x_j}}}, \label{subeq:3}
    \end{equation}
    \end{subequations}
    where $\rho$ and $p$ represent the density and pressure, respectively, $u_i$ is the flow velocity component, $\tau_{ij}$ is the Reynolds stress tensor, $\sigma_{ij}$ represents the viscous stress, which is calculated by:
    \begin{equation}
    \label{eq:RStress}
       {\sigma _{ij}} = \mu \left( {\frac{{\partial {u_i}}}{{\partial {x_j}}} + \frac{{\partial {u_j}}}{{\partial {x_i}}} - \frac{2}{3}\frac{{\partial {u_k}}}{{\partial {x_k}}}{\delta _{ij}}} \right) \text{,} 
    \end{equation}
    and $\mu$ is the viscosity.
    In the energy equation~\eqref{subeq:3}, $E$ and $H$ represent the total energy and total enthalpy, respectively, ${q_j^{\left( \text{t} \right)}}$ is the turbulent heat flux, and ${q_j}$ is the laminar heat flux. For details on terms in equation~\eqref{eq:NSEquations}, 
    The Reynolds stress $\tau_{ij}$ is obtained using the SSG/LRR--$\omega$ Reynolds stress model (RSM)~\cite{cecora2015differential,Liu2020RSM-IDDES}.
    The turbulent heat flux is modeled based on the gradient diffusion hypothesis~\cite{wilcox1998turbulence,blazek2015computational}.
    
    An in-house computational fluid dynamics (CFD) solver \cite{liu2019numerical, liu2018dynamic, Liu2020RSM-IDDES, wang2021iddes} is utilized to solve the RANS equations~\eqref{eq:NSEquations}. 
    The governing equations are discretized via a cell-centered finite volume approach on unstructured hybrid meshes consisting of hexahedrons, prisms, tetrahedrons, and pyramids. 
    Convective flux terms are computed using second-order Roe discretization schemes~\cite{roe1986characteristic}, while viscous flux terms are determined through a reconstructed central scheme. 
    For unsteady flow simulations, a second-order accurate implicit dual time-stepping method is employed. 
    To counteract the adverse effects of low-quality grids on solution stability and convergence, an adaptive local time-stepping technique is used. 
    All computations are executed with double precision on a high-performance computing (HPC) platform, with the CFD code parallelized through a domain decomposition strategy utilizing the message passing interface (MPI) protocol. 
    Nonblocking communications are employed to overlap computational tasks with communication, enhancing potential performance gains. The verification and validation (V\&V) of the CFD Solver are presented in~\ref{sec:AppendixA} and~\ref{sec:AppendixC}, respectively.

\subsection{Flow control strategies}

\begin{figure}[!htb]
    \centering
    \centering
    \subfloat[Circular cylinder]{
    \includegraphics[height=0.26\textwidth]{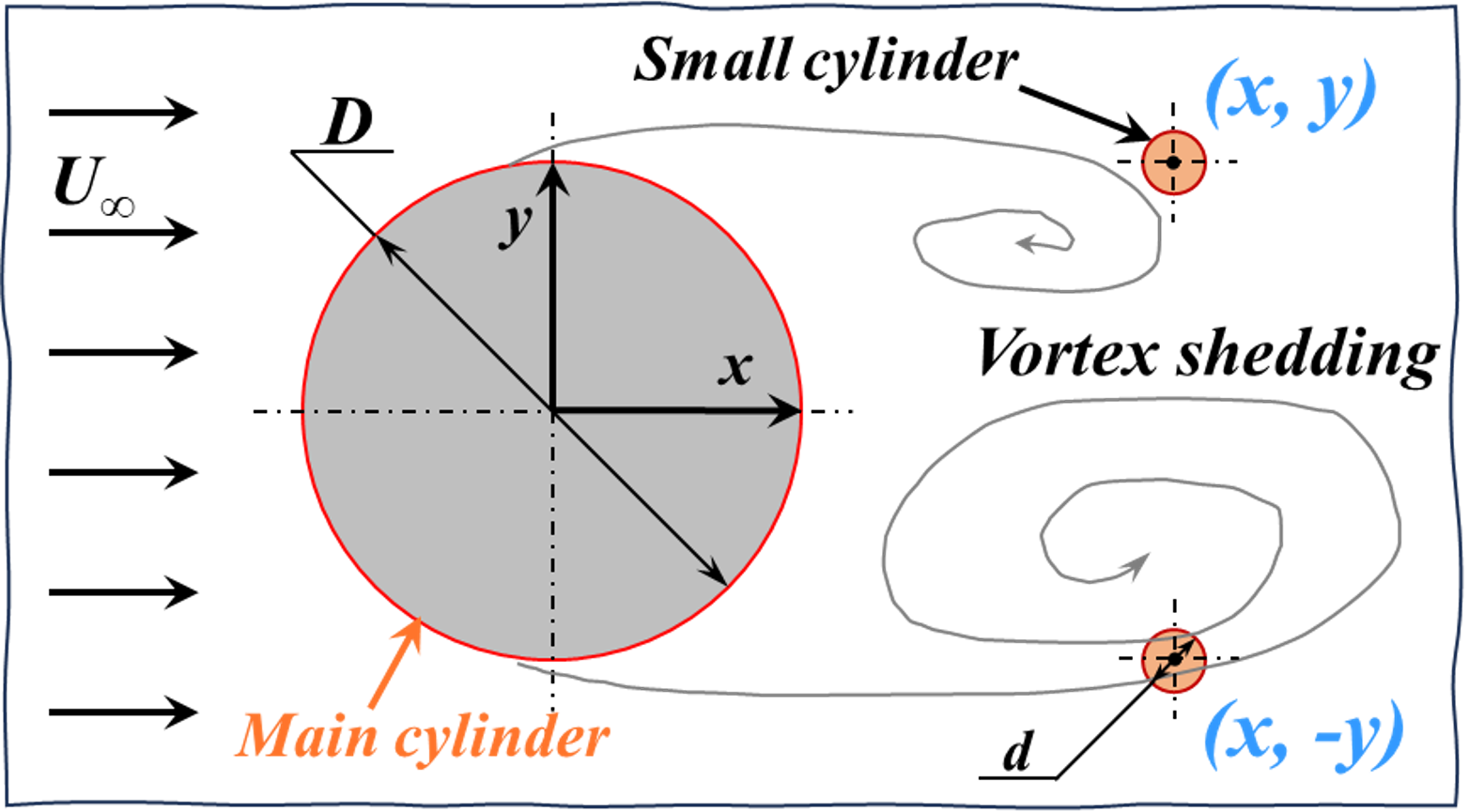} }
    \hspace{2mm}
    \subfloat[NACA 0012 airfoil]{
    \includegraphics[height=0.26\textwidth]{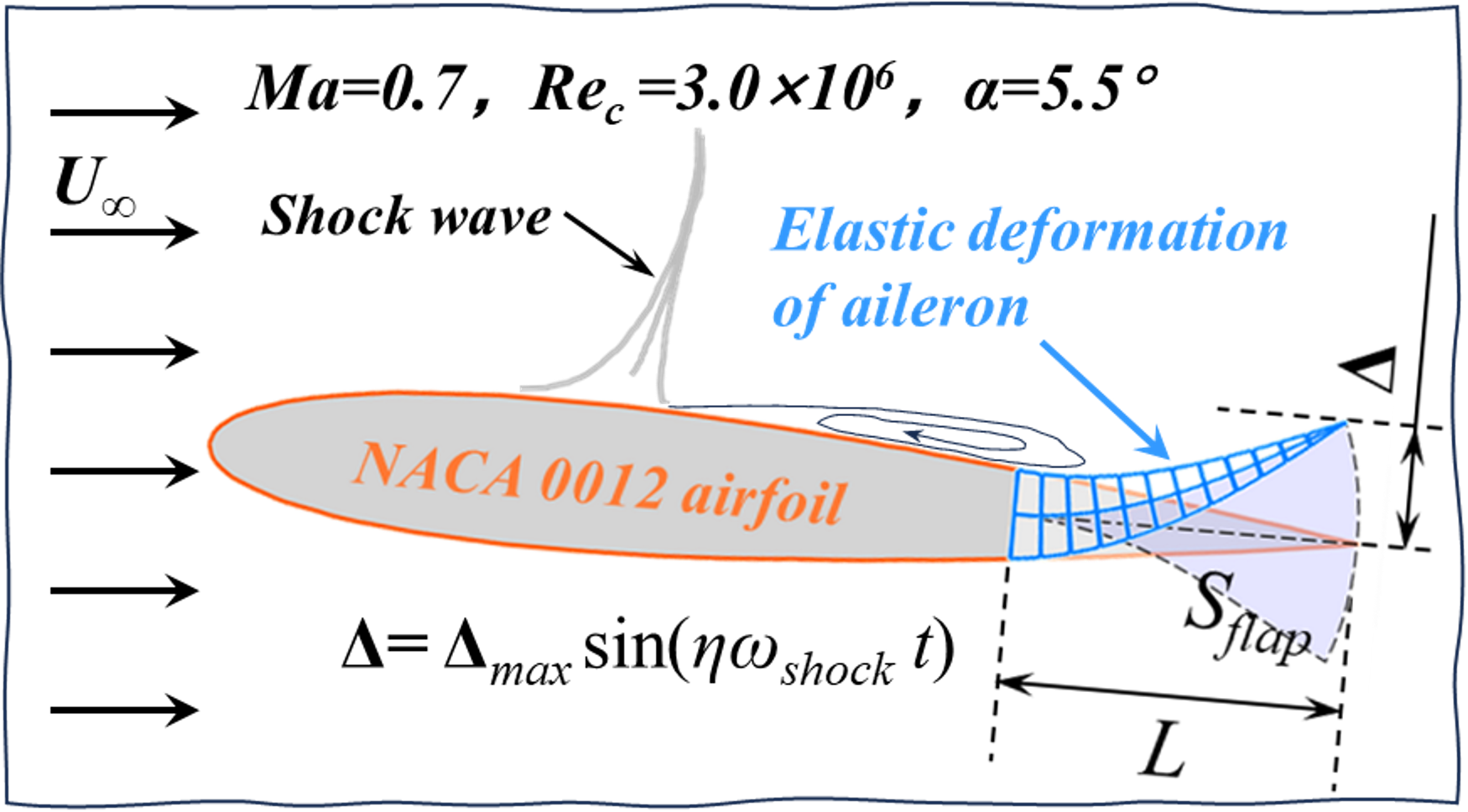} }
    \caption{Schematic plots of the flow control techniques:  (a) placement of small cylinders and (b) flapping of a compliant aileron.}
    \label{fig:sketchControl}
\end{figure}

In this work, we employ two different control techniques to show the capability of the ensemble-based method for flow control optimization.
One is the placement of small cylinders, and the other is the flapping of a compliant aileron.
The schematic view of these control problems is shown in Figure~\ref{fig:sketchControl}.
The details of the two control problems are briefly illustrated in the following.

The placement of small cylinders is used as the actuator to suppress the vibration of the flow over a circular cylinder.
This actuator is imposed in a passive manner, i.e., fixed optimal positions without external energy inputs.
In this control problem, as shown in Figure~\ref{fig:CylinderMesh}(a), three cylinders are arranged in an isosceles triangular pattern, and the position of the small cylinders~$(x, \pm y)$ is optimized to suppress vortex shedding behind the main cylinder.
The diameter of the main cylinder is $D$, and the diameter of the two small cylinders is $d$. The ratio of the diameters of the small and main cylinders is $d/D=0.125$.

The compliant aileron is used as the actuator to suppress the shock buffeting over the NACA 0012 airfoil.
The compliant aileron is an innovative active flow control technique, often referred to as the adaptive compliant trailing edge~\cite{herrera2015aeroelastic}. 
By controlling the deformation of the trailing edge, shock buffeting over an airfoil can be altered effectively. 
This actuator is imposed in both open-loop and closed-loop ways.
For the open loop control, the compliant aileron flaps at a fixed frequency and magnitude.
For the closed-loop control, the aileron movement is adjusted dynamically with the ensemble method based on real-time feedback.
Figure~\ref{fig:sketchControl}(b) shows the NACA 0012 airfoil equipped with a compliant aileron, where the deformation of the aileron follows the deflection curve equation of a cantilever beam~\cite{gere1997mechanics}.
The trailing edge of the airfoil is forced to oscillate with a sinusoidal function, $\Delta=\Delta_{max} \sin \left( \omega_{flap} t\right) =\Delta_{max} \sin \left( \eta \omega_{shock} t \right)$. 
Here, $\Delta_{max}$ and $\omega_{flap}$ are the amplitude and the frequency of the oscillating flap, respectively, and $\omega_{shock}$ is the frequency of the oscillating shock.
$S_{flap}$ ($S_{flap}=3\Delta_{max}L/8$) represents the area swept by the camber of aileron during one cycle of deformation, and $S_{flap}$ should be minimized to reduce energy input in practical applications.
In this control problem, the frequency~$\eta$ and magnitude~$\Delta_{max}$ of the aileron deformation, and the aileron length~$L$ are optimized to alleviate the shock buffeting.

\subsection{Ensemble Kalman method for flow control optimization}
\label{sec:2.2-ML_FlowControl}

We employ the ensemble Kalman method to optimize the control parameters of different flow control strategies.
The method is a statistical inference approach based on Monte Carlo sampling, which has been widely used in various applications~\cite{zhang2020evaluation,schneider2022ensemble,zhang2022acoustic}.
It randomly samples the uncertain parameters and uses the statistics of the control parameters and model predictions to estimate the gradient and Hessian of the cost function.
The cost function can be written as
\begin{equation}
\label{eq:ObjFunctions}
    J= \| w^{n+1} - w^n  \|_\mathsf{P} + \| \mathcal{H}[w^{n+1}] \|_\mathsf{R},
\end{equation}
    where $w$ is the control parameters to be optimized such as the position of actuators, $n$ is the iteration index, $\mathcal{H}$ is the model operator that maps the control parameters to the objective quantities, and $\mathsf{P}$ and $\mathsf{R}$ are the weight matrix.
    The weight~$\mathsf{P}$ and $\mathsf{R}$ can affect the optimization process significantly.
    Specifically, small sample variance~$\mathsf{P}$ would limit the possible solutions in the vicinity of the initial control parameters.
    Too large values of $\mathsf{P}$ can lead to large sample variance and update step length, which often causes optimization divergence for nonlinear problems as the linearization assumption does not hold in the ensemble-based gradient~\cite{zhang2024large}.
    Similarly, too small values of $\mathsf{R}$ can also result in large update step length and further optimization divergence.
    Too large values would lead to the ignorance of the objective term in the cost function and the convergence to the initial value.
    In this work, the standard deviation for~$\mathsf{P}$ is set as~$0.5$, and the standard deviation for~$\mathsf{R}$ is set as $0.1$, which is determined based on our sensitivity study.
    In order to mitigate the flow-induced vibration, we regard the standard deviation of the drag and lift force as the objective quantity~$\mathcal{H}[w]^{n+1}$.
    Also, one can include other quantities such as the mean drag within the objective to reduce the drag force simultaneously.
    The weight matrix~$\mathsf{P}$ is estimated based on the ensemble of the realizations $\{w_m\}_{m=1}^M$ as
\begin{equation}
\begin{aligned}
	\bar{W} &= \frac{1}{M} \sum_{m=1}^M w_{m} \text{,} \\
	\mathsf{P} &= \frac{1}{M-1} (W - \bar{W})(W-\bar{W})^\top \text{,}
\end{aligned}
\label{eq:cov}
\end{equation}
    where $M$ is the sample size and $m$ is the sample index.
    Large ensemble sizes can reduce the sampling errors and improve the robustness of the ensemble Kalman method. 
    However, unsteady CFD applications are often computationally time-consuming, and hence using large ensemble sizes will be impractical.
    Here we choose $50$ samples to achieve a balance between the inversion efficiency and robustness, which is commonly used in literature for ensemble Kalman inversion in CFD applications~\cite{zhang2020evaluation,zhang2022acoustic}.
    
    The sampling error and covariance collapse are important issues frequently encountered in the ensemble Kalman method.
    The sampling error can cause underestimation of the sample variance due to limited sample sizes~\cite{houtekamer1998data,van1999comment}.
    Moreover, it would provide spurious correlation~\cite{carrassi2018data}, leading to incorrect update directions and further optimization divergence.
    One can increase the sample size to alleviate this issue but at significant computational costs, particularly for unsteady CFD applications.
    Alternatively, the correlation-based localization technique~\cite{kelly2014well,luo2018correlation} may be introduced to alleviate the issues of sampling errors.
    On the other hand, the covariance collapse can severely affect the optimization convergence.
    That is, the samples may converge to the variance-minimizing solution due to the sample collapses~\cite{evensen2018analysis}, instead of the minima of the cost function.
    The inflation techniques~\cite{luo2024ensemble} can be introduced to address this issue.
    The localization and inflation techniques are worthy of further investigation to improve the performance of the ensemble method in solving optimization problems.
    Note that using a fixed covariance~$\mathsf{P}$ can avoid the sample collapse issue, while it is difficult to choose an appropriate value to have good convergence speed and optimized results.
    Small values of $\mathsf{P}$ lead to slow convergence speed, while large values may break the linear assumption in the ensemble-based gradient and cause the optimization divergence for nonlinear problems.

The ensemble Kalman method can update the control parameters based on the Gauss-Newton method, which requires estimating the first and second-order derivatives of the cost function.
The method uses the statistics of these samples to estimate the derivative information~\cite{luo2015iterative,zhang2022ensemble}.
At the $n$~th iteration, each sample~$w_m$ can be updated based on
\begin{equation}
    w_m^{n+1} = w_m^n - \mathsf{PH}^\top (\mathsf{HPH}^\top + \mathsf{R})^{-1} \mathcal{H}[w_m^n] \text{,}
    \label{eq:enkf}
\end{equation}
where $\mathsf{H}$ is the tangent linear model operator.
In practice, the operator~$\mathsf{H}$ is not needed to compute explicitly by reformulating $\mathsf{PH}^\top=\text{cov}(w, \mathcal{H}[w])$ and $\mathsf{HPH}^\top=\text{cov}(\mathcal{H}[w], \mathcal{H}[w])$, where $\text{cov}$ indicates the sample covariance of the two random variables.
Note that we add random perturbations into the objective value~$\mathcal{H}[w]$ based on the analysis scheme of stochastic ensemble Kalman method, which can alleviate the variance underestimation caused by the limited sample sizes~\cite{burgers1998analysis}.
The readers are referred to Ref.~\cite{luo2015iterative,zhang2020evaluation} for details of the ensemble-based optimization framework.

\begin{figure}[!htb]
    \centering
    \includegraphics[width=0.93\textwidth]{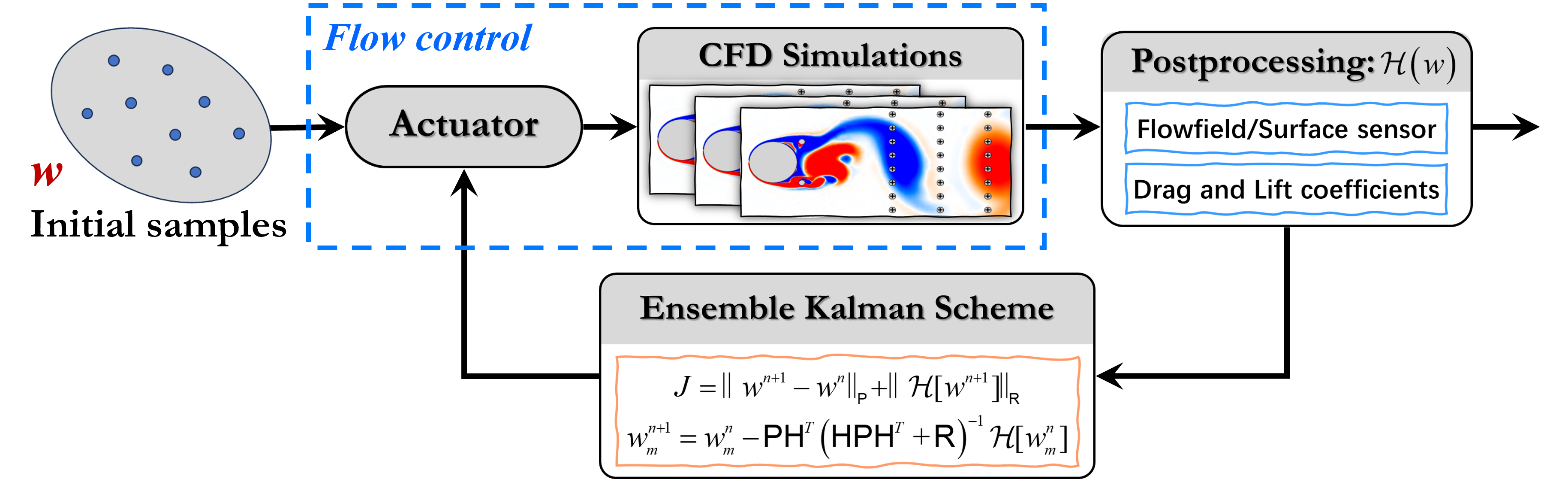} 
    \caption{Schematic of the ensemble-based Kalman method for flow control.}
    \label{fig:schematic}
\end{figure}

The schematic of the ensemble Kalman method for flow control is shown in Fig.~\ref{fig:schematic}.
Given the initial parameters~$w^0$ and the initial variance,  an ensemble of control parameters can be drawn randomly based on Gaussian distribution.
Each sample of the control parameter is used to update the control strategy.
Under the new control setting, the flow prediction~$\mathcal{H}[w_m]$ is obtained by solving the unsteady RANS equations.
Further, the model predictions are analyzed to update the control parameters based on the ensemble Kalman method.
The public DAFI code~\cite{strofer2021dafi} is used to implement the ensemble-based flow control method in this work.

The detailed optimization procedure is formulated as follows.
\begin{enumerate}[label=(\alph*)]  
    \item Initial sampling. 
    The samples of control parameters are drawn randomly around initial values based on a prescribed normal distribution.
    \item RANS simulation.
    For each sample of control parameters, the URANS simulation is performed under the corresponding control strategy, e.g., the trailing edge flapping.
    \item Evaluation of flow-induced vibration:
    Objective quantities such as the standard deviation of lift are obtained based on the simulated flow field. 
    It is used to evaluate the vibration on the structures for each sample of control parameters.
    \item Kalman update:
    The sample statistics are computed based on Eq.~\eqref{eq:cov}. 
    Further, the control parameters are updated by analyzing the evaluated objective quantity based on the ensemble Kalman method.
    Return to the step (b) until the maximum iteration is reached.
\end{enumerate}

\section{Numerical results}
\label{sec:3-NumericalResults}

    We present two flow cases to demonstrate the capability of the ensemble method for flow control optimization, including the vortex shedding from a circular cylinder and the transonic buffeting around the NACA 0012 airfoil. 
    Three different control tasks are applied, i.e., passive control of vortex shedding using small cylinders, open-loop active control of buffeting over an airfoil with a compliant aileron, and closed-loop active control of buffeting through dynamic adjustments of the aileron based on real-time feedback in lift and drag coefficients.
    The setup of each case is listed in Table~\ref{tab:DetailsSet} in terms of flow conditions, control parameters, objective functions, and so on.
    In Table~\ref{tab:DetailsSet}, the lift and drag forces are normalized to be the lift and drag coefficient ($C_l, C_d$) and calculated by
    \begin{equation}
    \left\{ \begin{array}{l}
    {C_l} = \frac{{{F_l}}}{{{\frac{1}{2}}\rho_\infty U_\infty ^2{L_{ref}}}} \\
    {C_d} = \frac{{{F_d}}}{{{\frac{1}{2}}\rho_\infty U_\infty ^2{L_{ref}}}}
    \end{array} \right. \text{,}
    \label{eq:ClAndCd}
    \end{equation}
    where ${F_l}$ and ${F_d}$ are lift and drag force, $\rho_\infty$ and $U_{\infty}$ represents the density and velocity at infinity, and $L_{ref}$ represents the reference length.

\begin{table}[!htb]
    \caption{\label{tab:DetailsSet} Summary of test cases in terms of the flow conditions, objectives, controllers, and constraints.}
    \centering
    \begin{tabular}{p{2.8cm}p{6cm}p{3cm}p{3cm}}
    \hline
    \hline
                     &  \multirow{2}{*}{Passive flow control}  & \multicolumn{2}{c}{Active flow control}  \\
    \cline{3-4}
                     &                       & Open-loop      & Closed-loop \\
    \hline
    Cases            &  Circular cylinder    & \multicolumn{2}{l}{NACA 0012 airfoil} \\
    Flow conditions  & $Ma=0.03, Re_D=3900$  & \multicolumn{2}{l}{$Ma=0.7, \alpha=5.5^\circ, Re_C=3.0 \times {10^6}$}  \\
    Flow features    & Vortex shedding after a bluff body & \multicolumn{2}{l}{ Transonic shock buffeting over an airfoil} \\
    Objective        & 1. averaged $C_d$; 2. vibration of $C_d$; 3. $u_{\text{rms}}$ in the wake. &  \multicolumn{2}{p{6cm}}{ 1. vibration of $C_d$; 2. vibration of $C_l$.} \\
    Control parameters  & Positions of small cylinders & \multicolumn{2}{p{6cm}}{Length of aileron; amplitude and frequency of aileron flapping} \\
    Constraint       & No overlapping between cylinders  &  \multicolumn{2}{l}{Limited aileron length $L \leq 0.25C$ $^a$} \\
    \hline
    \hline
    \multicolumn{4}{l}{ $^a$ $C$ represents the chord length of the airfoil. }
    \end{tabular}
\end{table}

    For the circular cylinder flow case, the incoming flow conditions are $Ma=0.03$, and the Reynolds number based on the diameter of the cylinder is $Re_D=3900$.
    In this case, our objective is to suppress lift vibrations, drag force, and turbulence intensity in the wake. 
    The objective function is accordingly designed to reduce the mean drag coefficient ($\langle{C_d}\rangle _\text{mean}$), the drag vibrations ($\langle{C_d}\rangle _{\text{std}}$), and the fluctuating velocity ($U_{\text{rms}}$) in the wake,
    which can be expressed as $\mathcal{H}[ {{w}}] = {\left[ {\langle{C_d}\rangle _\text{mean},\langle{C_d}\rangle _{\text{std}}}, U_{\text{rms}} \right]^\top}$.
    To prevent overlapping of the cylinders, the distance between the cylinder centers must be larger than the sum of their radii.
    In the NACA 0012 airfoil case, the incoming flow conditions are $Ma=0.7$ with an angle of attack of $\alpha=5.5$, and the Reynolds number based on the chord length ($C$) is $Re_c= 3.0 \times 10^6$. 
    Under these conditions, the upper surface of the airfoil experiences shock buffeting, leading to severe vibrations in both lift and drag, which can have adverse effects on aircraft safety. 
    In this case, we aim to alleviate shock buffeting with a compliant aileron.
    With open-loop control, the objective function includes the standard deviation of the drag force ($\langle{C_d}\rangle _{\text{std}}$), and the standard deviation of the lift force ($\langle{C_l}\rangle _{\text{std}}$), which can be expressed as $\mathcal{H}[{{w}}] = {\left[ {\langle{C_d}\rangle _{\text{std}},\langle{C_l}\rangle_{\text{std}}} \right]^\top}$.
    With closed-loop control, the objective quantities~$\mathcal{H}[w]$ are derived from the lift and drag signals at the current time step in the unsteady simulation, which is illustrated in Section~\ref{sec:3.2-NACA0012} in detail.
    These objective quantities are normalized to alleviate the effects of their different magnitudes by multiplying a scaling factor on each quantity to have the same maximum value.
    To ensure practicality in engineering applications, the length $L$ of a compliant aileron is limited to less than $25\%$ of the airfoil chord length. 
    The constraints are imposed by bounding the violated samples at the threshold value in this work.
    
\subsection{Flow past a circular cylinder}
    \label{sec:3.1-Cylinder} 

\begin{figure}[!htb]
    \centering
    \centering
    \subfloat[Numerical setup]{
    \includegraphics[height=0.32\textwidth]{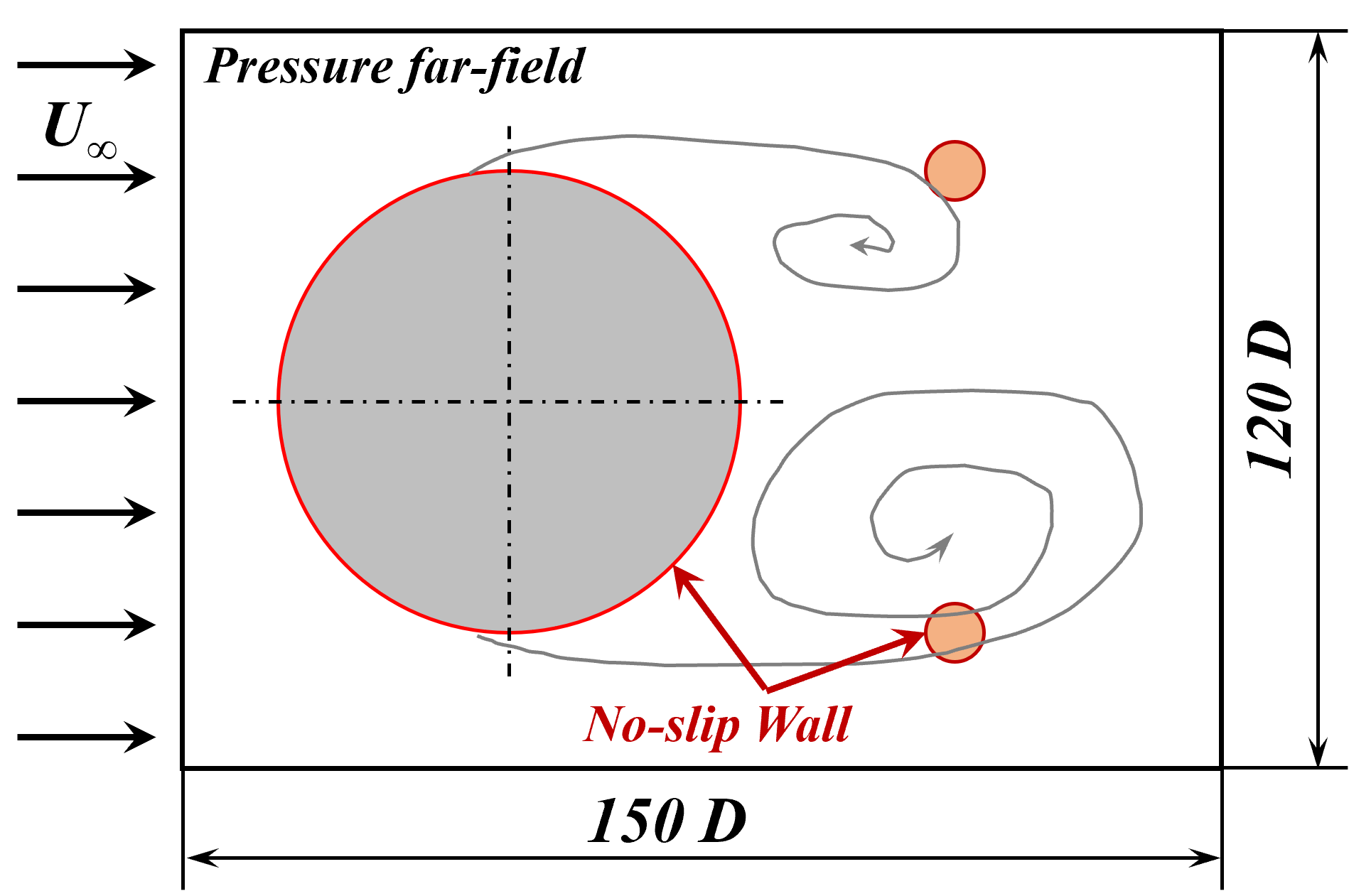} }
    \hspace{2mm}
    \subfloat[Computational mesh]{
    \includegraphics[height=0.32\textwidth]{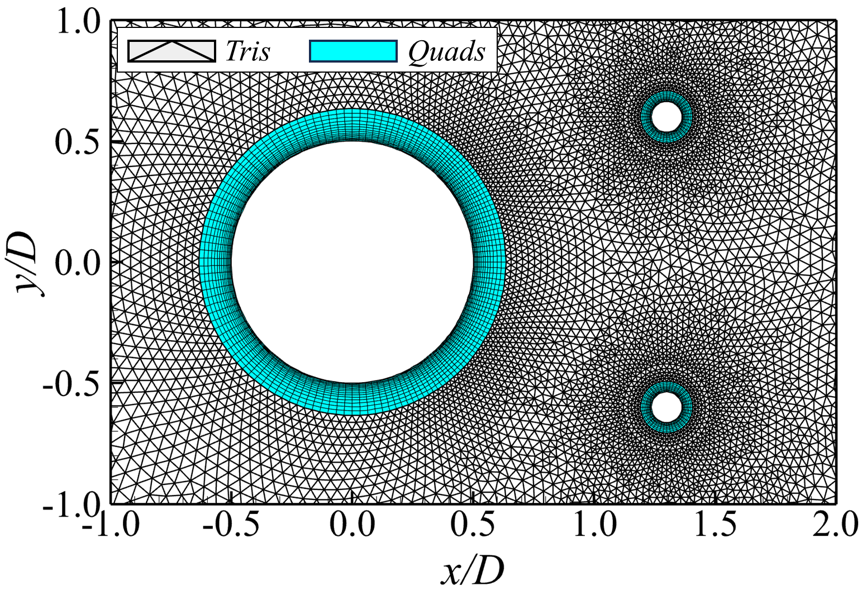} }
    \caption{Schematic plots of the computational model (a) and the hybrid unstructured mesh used for simulations of a circular cylinder (b). }
    \label{fig:CylinderMesh}
\end{figure}

    Figure~\ref{fig:CylinderMesh} (a) shows the computational domain for flows past the circular cylinder, which is a rectangle of length $150D$ and width $120D$, and the main cylinder is located at the center of the rectangle.
    An adiabatic no-slip wall condition is applied to the cylinder surface.
    A hybrid unstructured mesh is used to flow past the circular cylinder as shown in Fig.~\ref{fig:CylinderMesh}(b). 
    A body-fitted O-type structured mesh is generated in the nearby region of the wall, and triangular meshes are used in the remaining computational domain.
    These triangular meshes are regenerated using the advancing front method~\cite{wang2004mixed} when the positions of the two smaller cylinders are changed.
    The mesh is selected based on grid sensitivity tests as presented in~\ref{sec:AppendixA}. 
    The unsteady RANS simulations use $200$ time steps per cycle to capture the vortex shedding in the wake.

\begin{figure}[!htb]
    \centering
    \centering
    \subfloat[Locations of two smaller cylinders]{
    \includegraphics[height=0.33\textwidth]{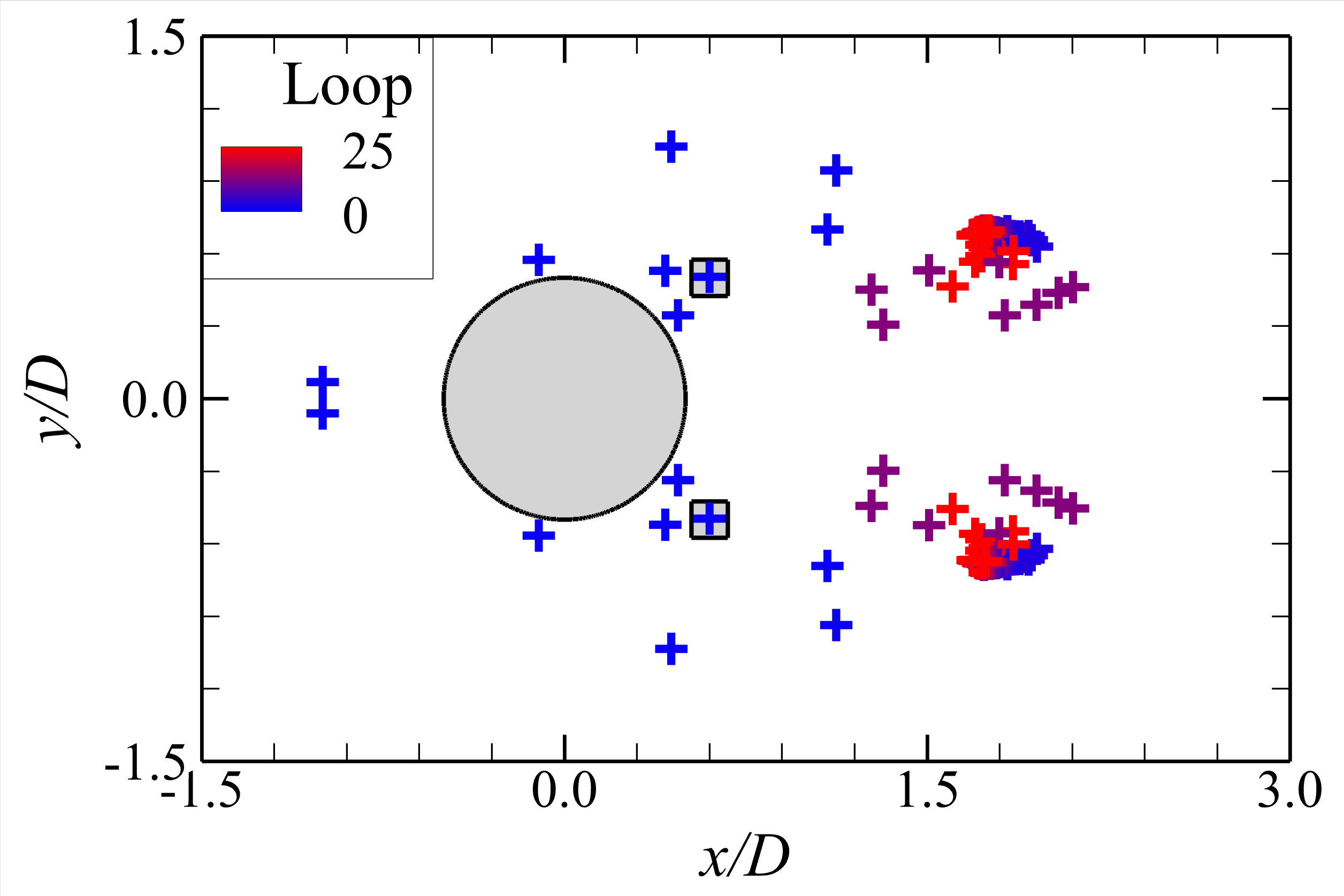} }
    \hspace{1mm}
    \subfloat[Convergence history of objective function]{
    \includegraphics[height=0.33\textwidth]{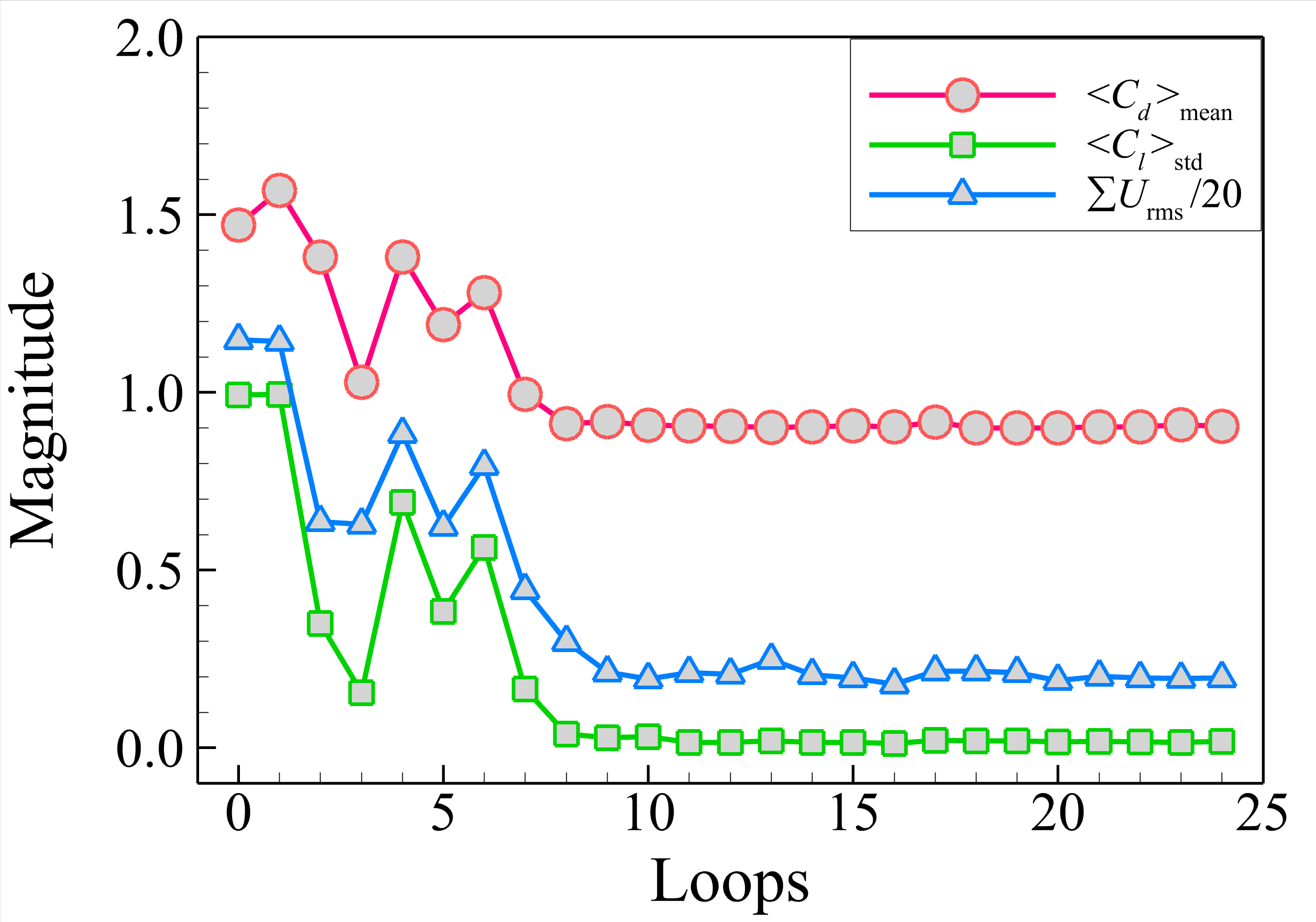} }
    \caption{ Convergence history of the optimization. (a) Locations of the smaller cylinders at each iteration; (b) Evolution of the $C_l$, $C_d$, and $U_{\text{rms}}$. }
    \label{fig:Convergence}
\end{figure}

    We employ the ensemble-based flow control approach to reduce drag and lift forces on a cylinder and suppress velocity fluctuations in the wake. 
    This involves arranging the positions of two smaller cylinders to alter the flow structure. 
    The two small cylinders are placed initially at $(0.6,0.5)$ and $(0.6,-0.5)$.
    Figure~\ref{fig:Convergence}(a) illustrates the cylinder locations at different optimization steps, with the ``$\boxplus$'' denoting the initial coordinates of the small cylinders. 
    The transition from blue to red for ``$+$'' indicates the iterative optimization process.
    As illustrated in the figure, the search space covers most of the near-wall region, extending to the upstream and downstream of the cylinder at both lateral sides. 
    After approximately ten iteration steps, the position of the small cylinders can converge to the optimal locations of $(1.752,\pm0.691)$. 
    Figure~\ref{fig:Convergence}(b) presents the evolution of the lift coefficient ($C_l$), drag coefficient ($C_d$), and velocity fluctuations ($U_{\text{rms}}$) with the cylinder positions altered. 
    Here, the $U_{\text{rms}}$ is the integral of the root mean square (RMS) velocity ($u_{\text{rms}}$) along the y-axis at one cross-section $x/D$, and it is defined as
    \begin{equation}
        {U_{\text{rms}}} = \int_{ - 3D}^{3D} {{u_{\text{rms}}}\left( y \right)dy}  \text{.}
        \label{eq:UrmsError}
    \end{equation}
    The integration zone is set as $y\in[-3D,3D]$ to cover the entire wake flow.
    The horizontal axis in Figure~\ref{fig:Convergence} (b) represents the number of iterations, and the vertical axis represents magnitudes.
    The $\sum{U_{\text{rms}}(x)}$ indicates the sum of velocity fluctuation $U_{\text{rms}}$ at six cross-sections, i.e., $x/D=2.5, 3.5, 4.5, 5.5, 6.5,$ and $7.5$.
    For scale consistency, $\sum{U_{\text{rms}}}$ is scaled down by a factor of $20$. 
    The results confirm the observation in Figure~\ref{fig:Convergence}(a), where the coefficients converge after ten iterations.

\begin{figure}[!htb]
    \centering
    \centering
    \subfloat[Vorticity field without control]{
    \includegraphics[width=0.465\textwidth]{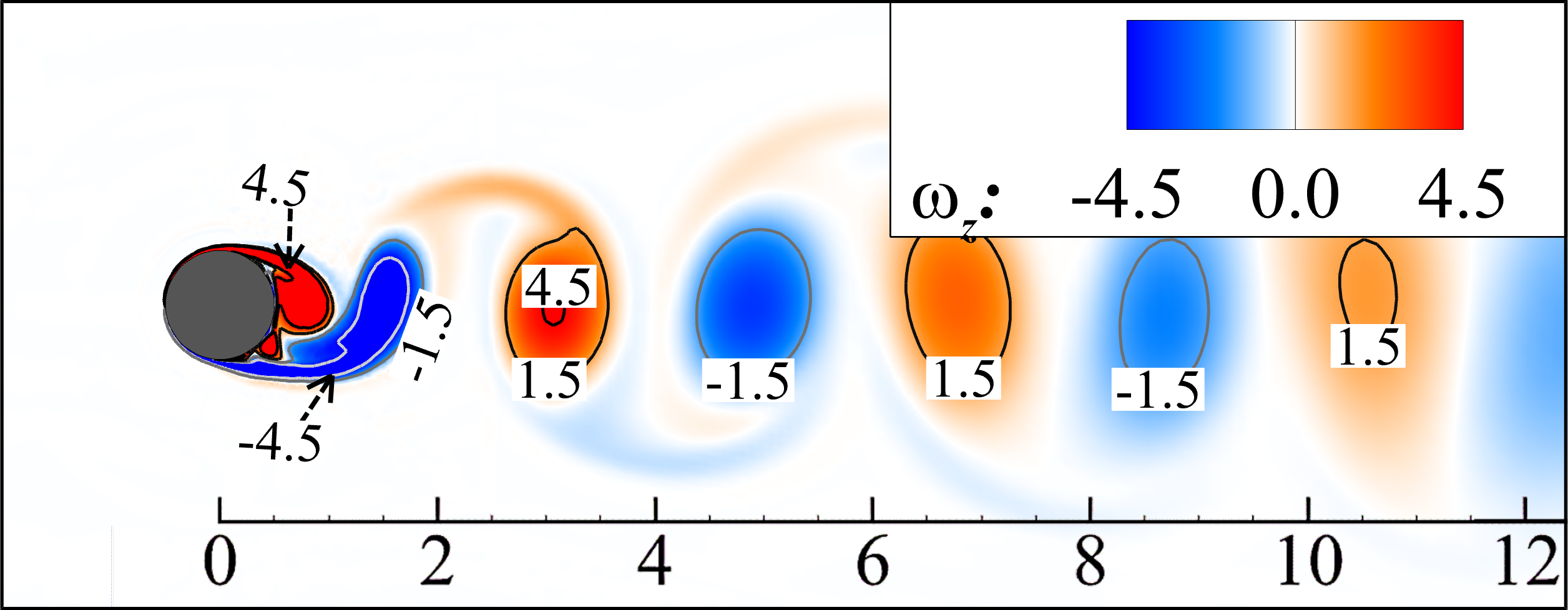} }
    \hspace{2mm}
    \subfloat[Vorticity field with control]{
    \includegraphics[width=0.465\textwidth]{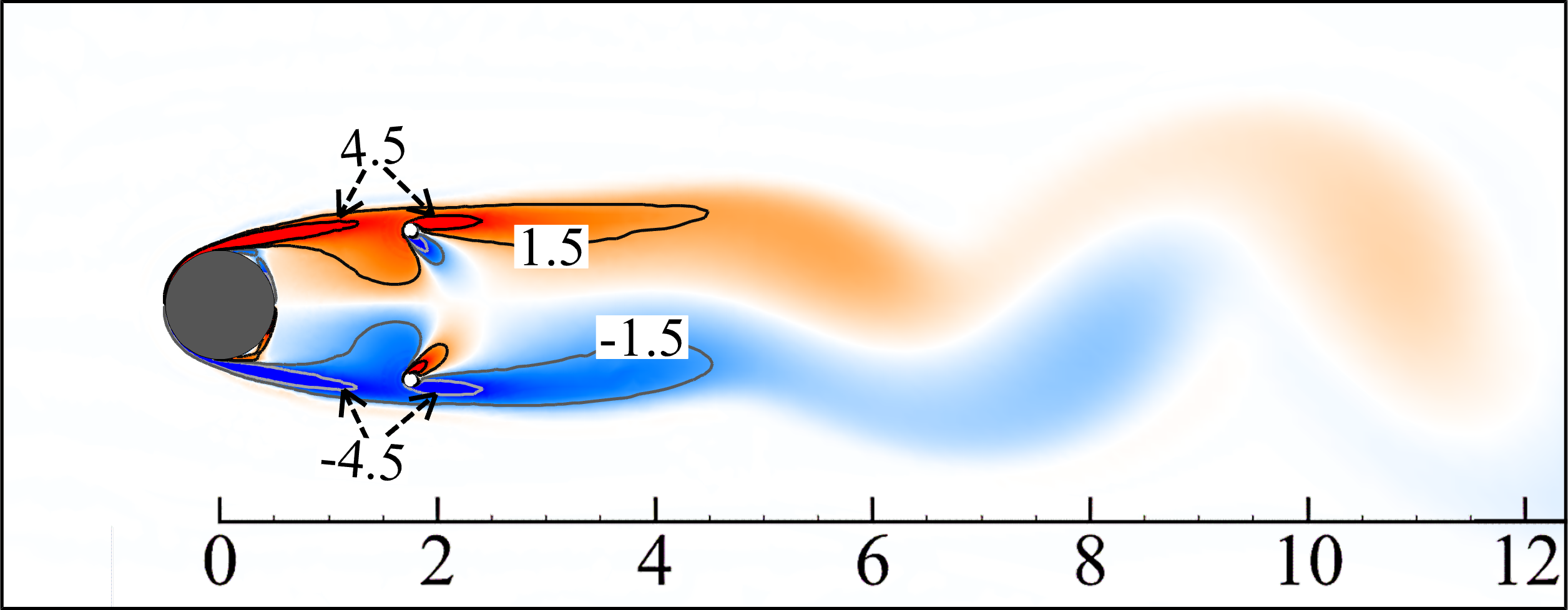} }  \\
    \subfloat[Velocity field without control]{
    \includegraphics[width=0.46\textwidth]{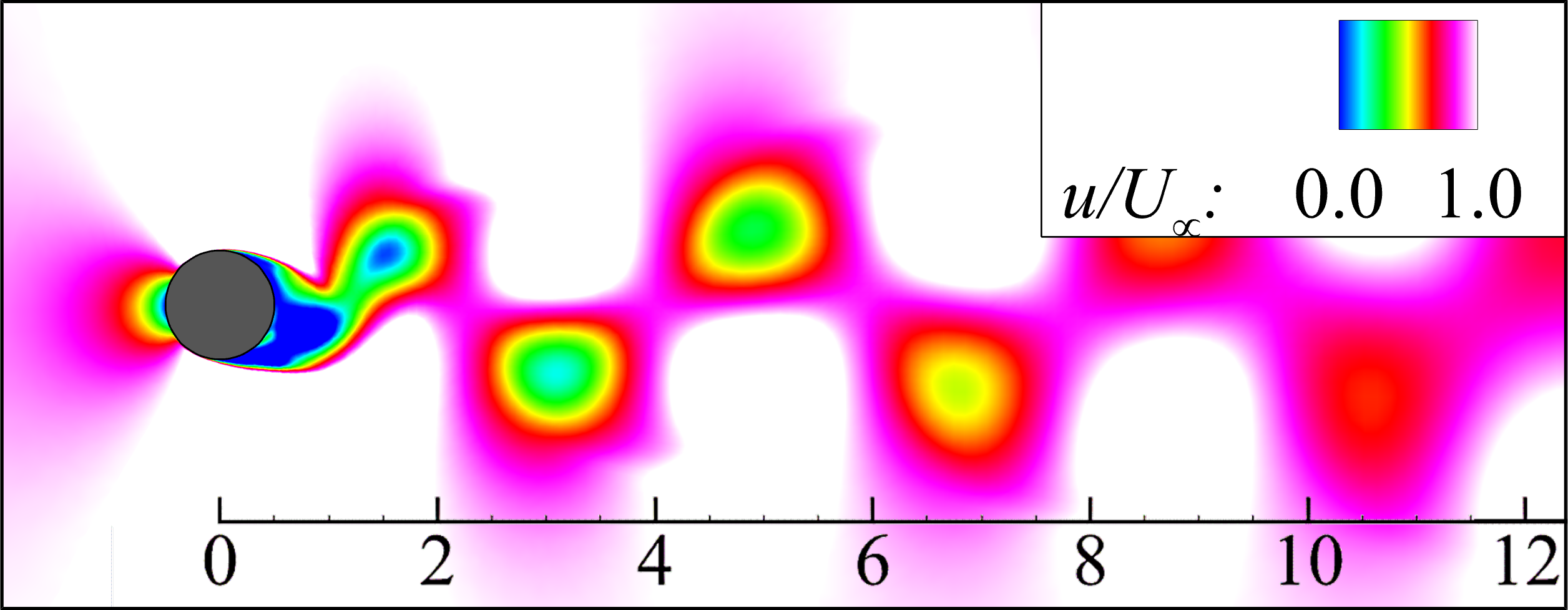} }
    \hspace{2mm}
    \subfloat[Velocity field with control]{
    \includegraphics[width=0.465\textwidth]{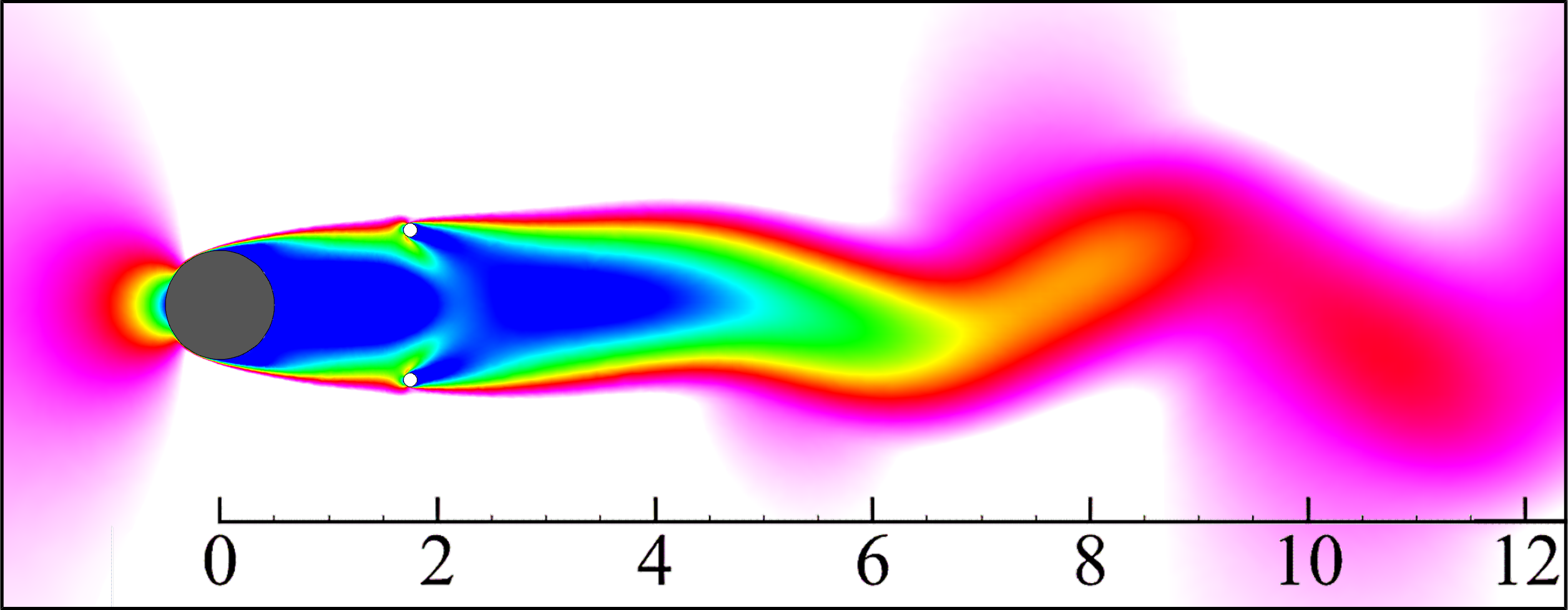} }
    \caption{Comparison of vorticity and velocity snapshots. Nondimensional vorticity $\boldsymbol{\omega}_z$ is obtained by ${\boldsymbol{\omega}_z} = \left( {\frac{{\partial v}}{{\partial x}} - \frac{{\partial u}}{{\partial y}}} \right)\frac{{{L_{ref}}}}{{{U_\infty }}}$. Each snapshot corresponds to the moment of maximum lift. }
    \label{fig:VorticityCompared}
\end{figure}

    The ensemble Kalman method optimizes the position of two small cylinders to effectively reduce both of the magnitude and frequency of vortex shedding behind the main cylinder.
    Figure~\ref{fig:VorticityCompared} presents instantaneous snapshots of vorticity ($\boldsymbol{\omega}_z$) and velocity ($u$) with and without the passive flow control.
    All field snapshots correspond to the typical instant of the maximum lift.
    It can be observed that the vortex shedding from the cylinder is significantly suppressed. 
    Specifically, the vorticity magnitude ($\boldsymbol{\omega}_z$) downstream of the main cylinder is reduced, and the location where vortex shedding occurs is shifted further downstream, compared with the uncontrolled case in Figure~\ref{fig:VorticityCompared} (a).

\begin{figure}[!htb]
    \centering
    \centering
    \includegraphics[width=0.46\textwidth]{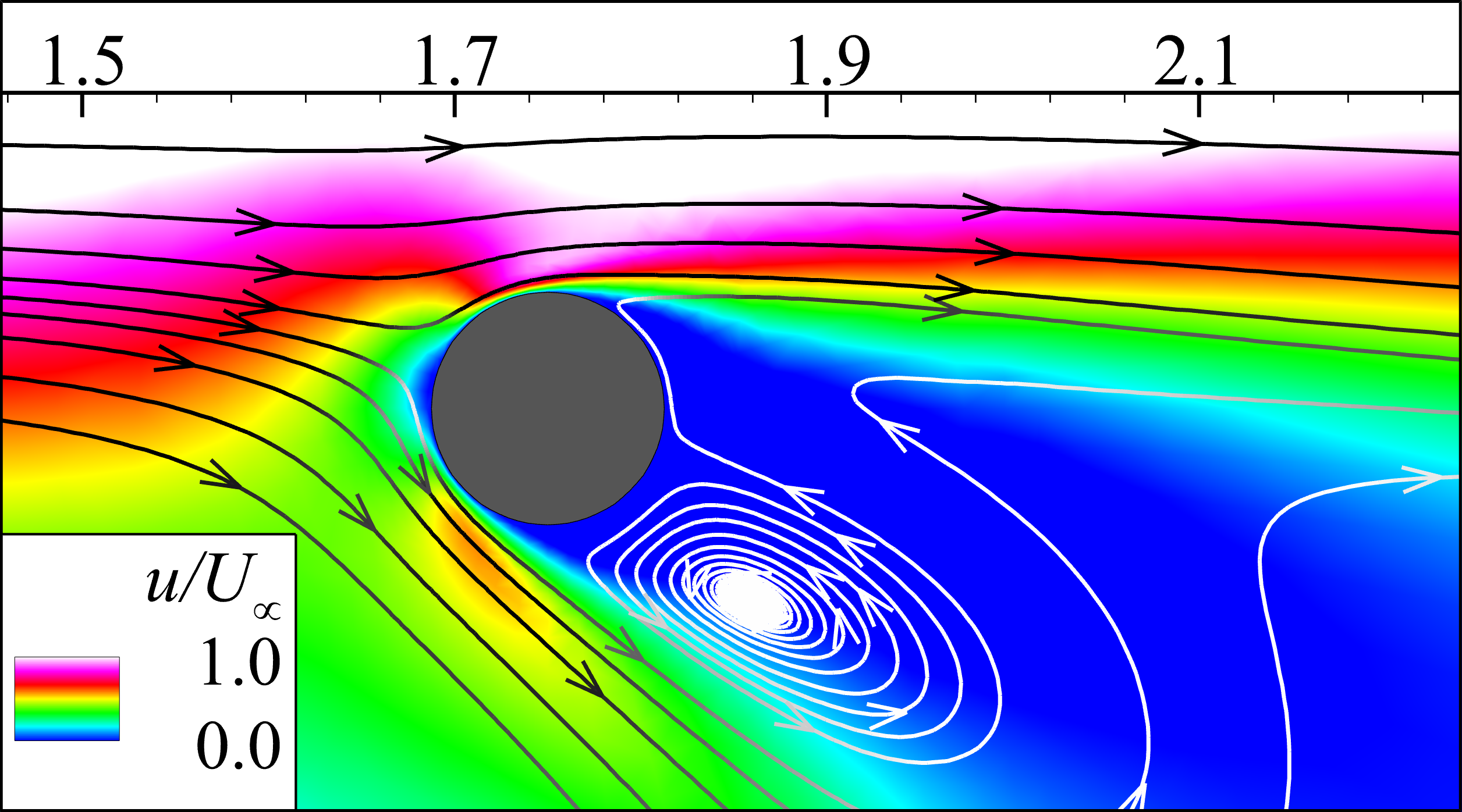}
    \caption{ Zoomed-in view of the instantaneous velocity field in the vicinity of the small cylinder at the moment of maximum lift.}
    \label{fig:VelocitySCylinder}
\end{figure}

    The suppressed vortex shedding is due to the formation of counter-rotating vortices behind the small circular cylinders in the controlled case. 
    These vortices effectively counteract the local vorticity, leading to a reduction in the amplitude of the vortices within the wake. 
    The vorticity fields, as shown in Figure~\ref{fig:VorticityCompared} (a) and (b), illustrate the differences between the uncontrolled and controlled cases. 
    In the uncontrolled case [Figure~\ref{fig:VorticityCompared} (a)], the vorticity field exhibits strong vortex shedding and high-amplitude vortices with $\left|{\boldsymbol{\omega}_z}\right| \geq 1.5$ extending up to $x/D \approx 11$. 
    In the controlled case [Figure~\ref{fig:VorticityCompared} (b)], these counter-rotating vortices weaken the vorticity strength and significantly compressing the area where $\left|{\boldsymbol{\omega}_z}\right| \geq 4.5$ to within the shear layers on each side of the main cylinder. 
    Additionally, the small cylinders diffuse the vorticity, making the vortex structures less concentrated and intense, with $\left|{\boldsymbol{\omega}_z}\right| \geq 1.5$ ending at $x/D \approx 4.15$.
    Figure~\ref{fig:VelocitySCylinder} provides a zoomed-in view of the instantaneous velocity field near the small cylinder. It is shown that the small cylinder is located at the shear layers of the main cylinder. The presence of small cylinders inhibits the motion of the shear layer developed from the main cylinder, resulting in an almost steady-state flow. Moreover, a quasi-steady flow regime exists downstream of the small cylinder, characterized by the vortex formation on only one side due to the influence of a non-uniform inflow. 

\begin{figure}[!htb]
    \centering    
    \subfloat[Flow without control]{
    \includegraphics[width=0.465\textwidth]{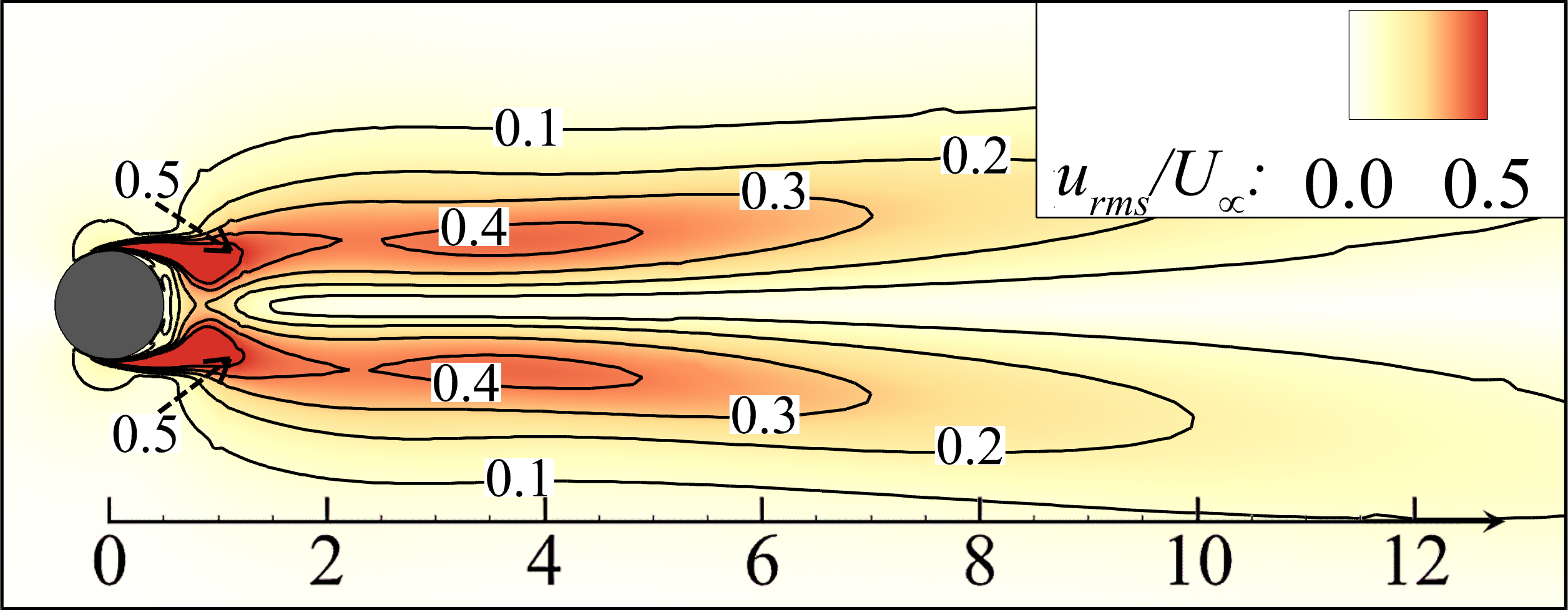} }
    \hspace{2mm}
    \subfloat[Flow with control]{
    \includegraphics[width=0.465\textwidth]{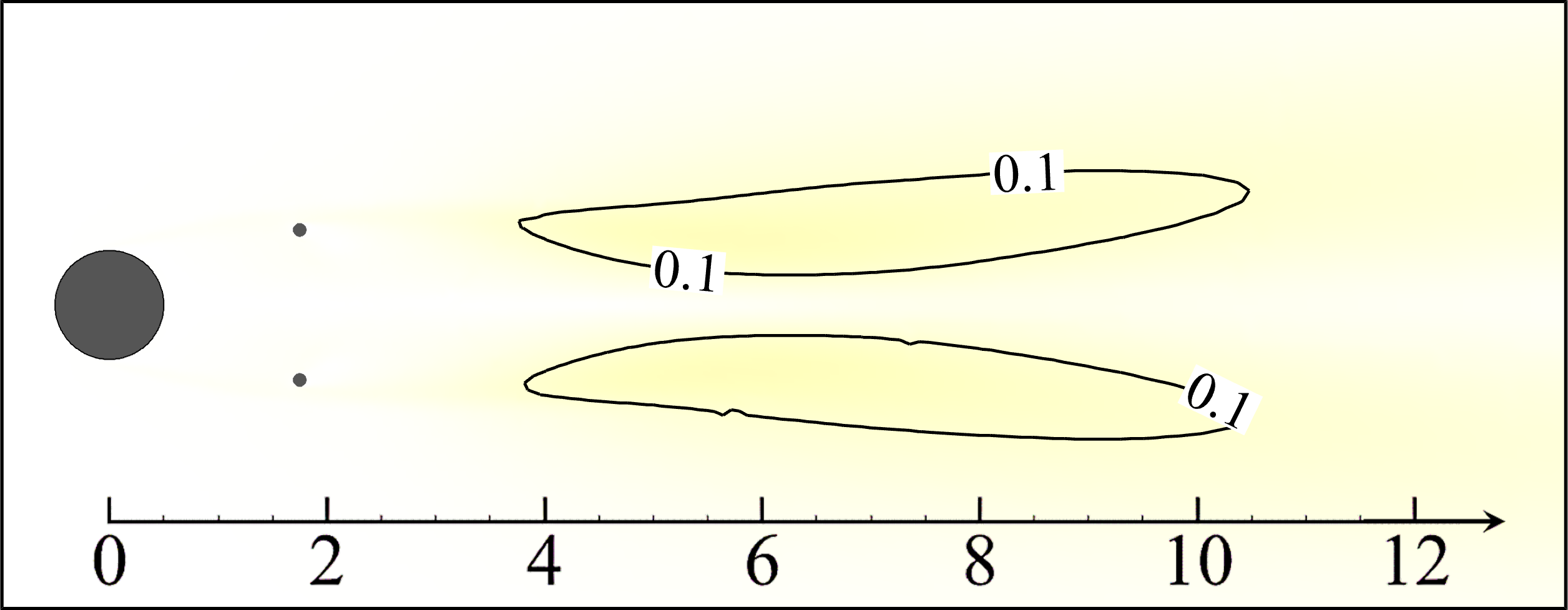} }
    \caption{Comparison of the RMS velocity ($u_{\text{rms}}$) contours in the wake for uncontrolled (a) and controlled (b) flows.}
    \label{fig:UrmsCompared}
\end{figure}

    The ensemble method is able to suppress the velocity fluctuations in the wake by optimizing the positions of the small cylinders.
    It is supported by Figure~\ref{fig:UrmsCompared}, where the root mean square (RMS) velocity ($u_{\text{rms}}$) contours between the two cases with and without flow control are compared. 
    The results clearly show a significant reduction of $u_{\text{rms}}$ within the wake region.
    Also, the areas with the highest flow fluctuations are shifted downstream.
    Specifically, without flow control, regions with significant fluctuations originate around the separation points and are amplified along the shear layers due to vortex shedding.  
    The maximum fluctuating velocity ($u_{\text{rms}} \geq 0.5$) occurs in two regions approximately $0.5D$ downstream of the cylinder, exhibiting a symmetrical distribution. 
    As the flow develops, the amplitude of $u_{\text{rms}}$ gradually decreases. 
    However, regions with $u_{\text{rms}} \geq 0.2$ extend to approximately $10D$ downstream of the cylinder.
    The passive flow control employing two small cylinders alters both the locations and magnitudes of these fluctuations. 
    As shown in Figure~\ref{fig:UrmsCompared} (b), the region with $u_{\text{rms}}\geq 0.2$ is eliminated, and the maximum $u_{\text{rms}}$ is reduced to 0.152. 
    Compared to the uncontrolled flow, the ensemble-based flow control significantly suppresses the region where $u_{\text{rms}} \geq 0.1$  into two narrow crescent-shaped regions that are symmetrically distributed at $x/D \in [4.0,10.5]$.

\begin{figure}[!htb]
    \centering
    \centering
    \includegraphics[width=0.93\textwidth]{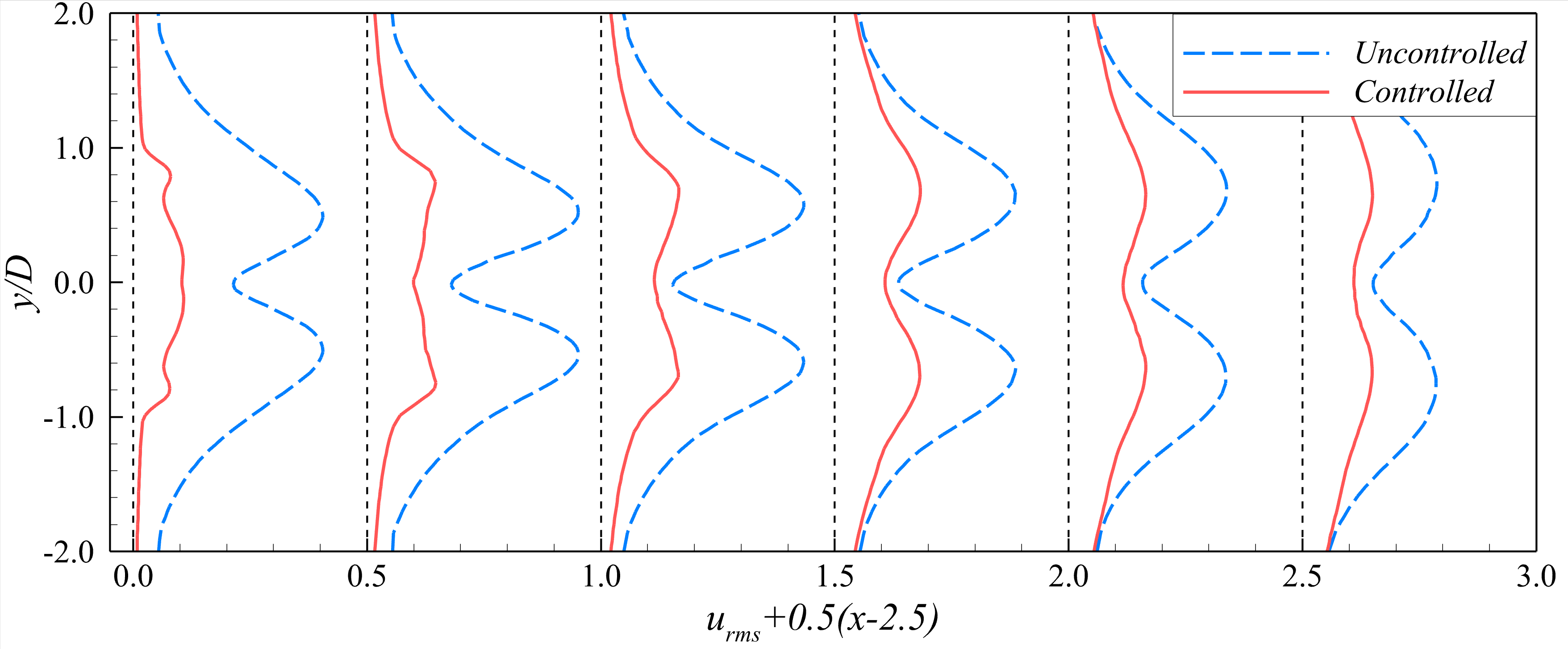}
    \caption{ Comparison of RMS-velocity plots in the wake at $x/D=2.5$, $x/D=3.5$, $x/D=4.5$, $x/D=5.5$, $x/D=6.5$, and $x/D=7.5$. }
    \label{fig:UrmsCdClCompare}
\end{figure}

    Figure~\ref{fig:UrmsCdClCompare} presents the profiles of $u_{\text{rms}}$ at five $x$-locations. 
    These profiles are located at $x/D=2.5$, $x/D=3.5$, $x/D=4.5$, $x/D=5.5$, $x/D=6.5$, and $x/D=7.5$, all of which are positioned downstream of the small cylinder. 
    Without flow control, the distribution of $u_{\text{rms}}$ is depicted by a blue solid line. 
    The $u_{\text{rms}}$ of the uncontrolled case at all sections exhibits two peaks that are symmetrically distributed on both sides of $y/D=0.0$. 
    These dual peaks are attributed to the vortex shedding downstream. 
    The introduction of two small cylinders, located at $(1.752,\pm0.691)$, can mitigate the peaks in the velocity fluctuations. 
    Also, the suppression effect is most pronounced at the position closest to the small cylinder.
    As the flow develops downstream, the control effect of the small cylinders slightly weakens, but it still reduces the amplitude of $u_{\text{rms}}$ by more than half.

    Table~\ref{tab:ErrorEstmtCylinder} summarizes the velocity fluctuations ($U_{\text{rms}}$) at five locations, the mean drag coefficient ($\langle{C_d}\rangle _\text{mean}$), and the standard deviation of lift coefficient ($\langle{C_l}\rangle _{\text{std}}$).
    We also define the relative change $\epsilon_\text{rel}$ between the prediction with and without flow control to quantify the improvement, which is based on 
    \begin{equation}
        \epsilon_\text{rel} = \frac{ q - q^\text{baseline} }{ q^\text{baseline} } \text{.}
    \label{eq:Rerror}
    \end{equation}
    It can be seen that there exists a significant reduction in $U_{\text{rms}}$ across all locations ($x/D=2.5, 3.5, 4.5, 5.5, 6.5, 7.5$), with relative changes spanning from $-79.971\%$ to $-50.037\%$. 
    This indicates a substantial decrease in velocity fluctuations under flow control, which effectively stabilizes the wake of the cylinder.
    Also, the ensemble-based method can optimize the flow control strategy to significantly reduce the mean drag and the vibration of lift as shown in Table~\ref{tab:ErrorEstmtCylinder}. 
    In the case under flow control, the time-averaged drag coefficient $\langle{C_d}\rangle _\text{mean}$ is reduced from its initial value of $1.088$ to $0.863$, resulting in a $20.680\%$ reduction. 
    And, the standard deviation of drag coefficient ($\langle{C_d}\rangle _{\text{std}}$) is reduced from $2.049 \times 10^{-2}$ to $9.068 \times 10^{-3}$, resulting in an $55.737\%$ reduction.

\begin{table}[!htb] 
    \caption{ \label{tab:ErrorEstmtCylinder} Summary of the relative change in the mean drag coefficient ($\langle{C_d}\rangle _\text{mean}$), the standard deviation of lift coefficient ($\langle{C_l}\rangle _{\text{std}}$), and the RMS velocity ($U_{\text{rms}}$). }
    \centering
    \begin{tabular}{ccccc}
    \hline
    \hline
                          &                & Uncontrolled     & Controlled   & Relative change ($\epsilon_\text{rel}$) $ ^a$\\
    \hline      
    \multirow{6}{*}{$U_{\text{rms}}$} & $x/D=2.5$ & $9.370 \times 10^{-1}$ & $1.877 \times 10^{-1}$ & -79.971\% \\
                               & $x/D=3.5$ & $9.864 \times 10^{-1}$ & $3.071 \times 10^{-1}$ & -68.865\% \\
                               & $x/D=4.5$ & $9.570 \times 10^{-1}$ & $3.668 \times 10^{-1}$ & -61.675\% \\
                               & $x/D=5.5$ & $9.049 \times 10^{-1}$ & $4.200 \times 10^{-1}$ & -53.584\% \\
                               & $x/D=6.5$ & $8.725 \times 10^{-1}$ & $4.197 \times 10^{-1}$ & -51.901\% \\
                               & $x/D=7.5$ & $8.076 \times 10^{-1}$ & $4.035 \times 10^{-1}$ & -50.037\% \\
    \hline
    \multicolumn{2}{c}{$\langle{C_d}\rangle _\text{mean}$}   & $1.088 $  & $0.863$ & -20.680\% \\
    \multicolumn{2}{c}{$\langle{C_d}\rangle _{\text{std}}$}  & $2.049\times 10^{-2}$ & $9.068 \times 10^{-3}$ & -55.737\% \\
    \hline
    \hline
    \multicolumn{5}{p{11cm}}{ $ ^a$ { The relative change ($\epsilon_\text{rel}$) is estimated based on Eq.~\eqref{eq:Rerror}, and negative sign indicates a reduction.} }
    \end{tabular}
\end{table}

\begin{figure}[!htb]
    \centering
    \centering
    \subfloat[$C_p$]{
    \includegraphics[width=0.46\textwidth]{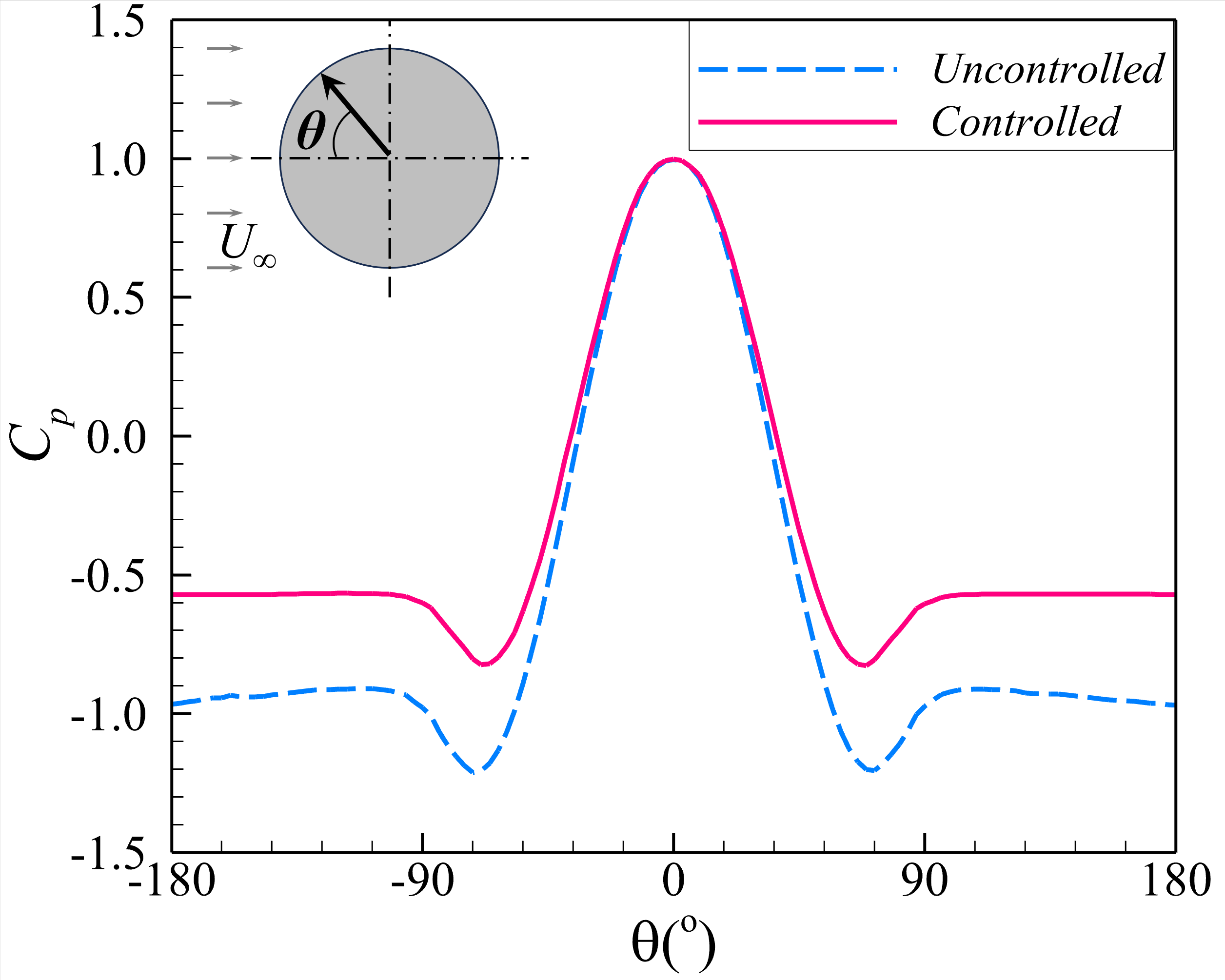} }
    \hspace{2mm}
    \subfloat[$C_f$]{
    \includegraphics[width=0.46\textwidth]{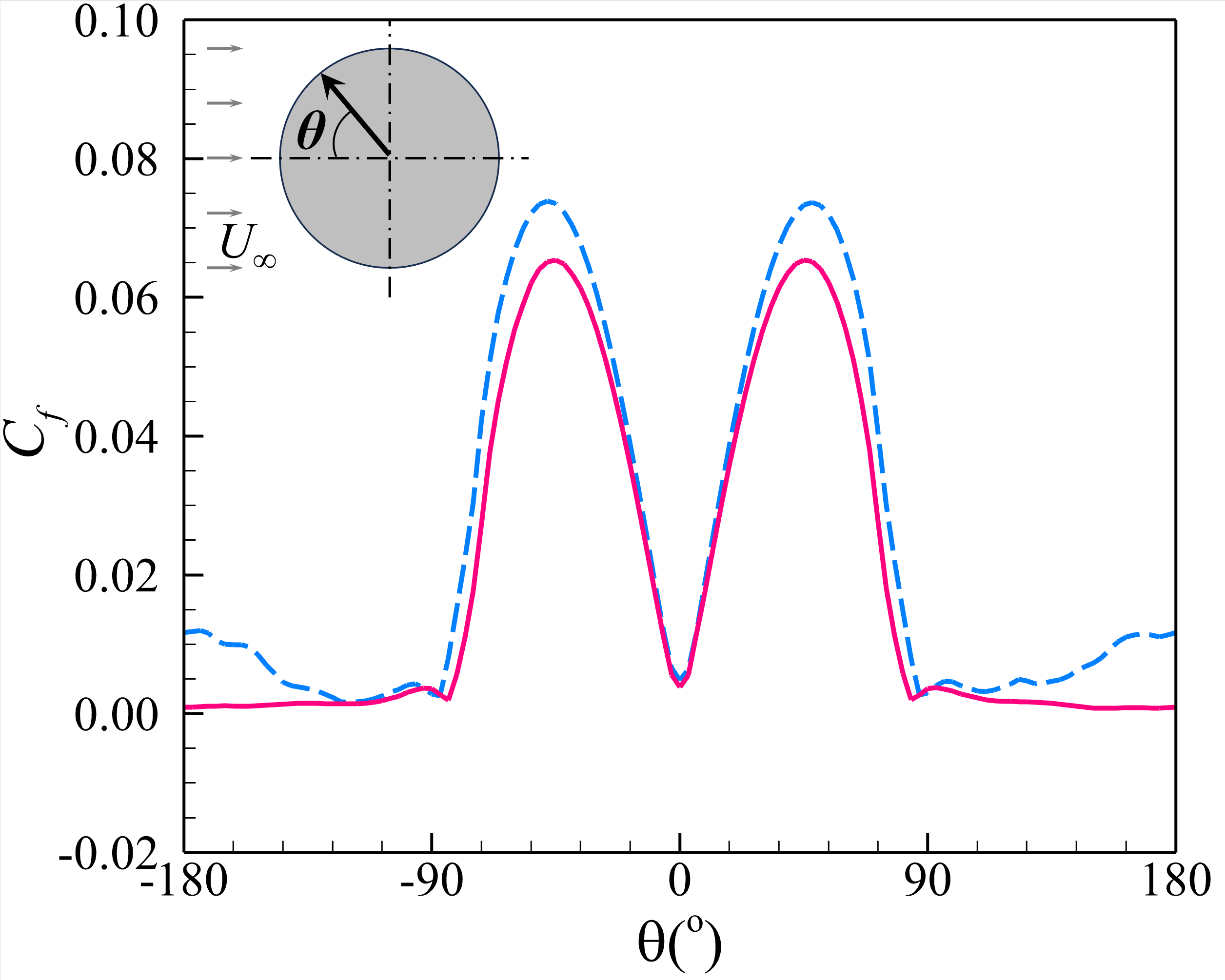} }
    \caption{Comparison of time-averaged $C_p$ (a) and $C_f$ (b) distributions between cases with and without flow control for the circular cylinder case.}
    \label{fig:CpCfCompared}
\end{figure}

    To investigate the reduction of mean drag and its associated vibrations, we further examine the pressure and friction on the main cylinder surface. 
    Figure~\ref{fig:CpCfCompared} presents the comparison of time-averaged $C_p$ and $C_f$ on the main cylinder surface for cases with and without flow control. 
    The comparison indicates that the mean drag reduction in the controlled case is mainly due to increased pressure on the leeward side of the main cylinder, as shown in Figure~\ref{fig:CpCfCompared} (a). 
    A separation zone can be observed downstream of the main cylinder in Figure~\ref{fig:VorticityCompared} (c) and (d), characterized by a negative horizontal component of velocity.
    The separated wake region is more extensive in the controlled case than in the uncontrolled one.
    This enlargement of the wake area leads to a reduced mean pressure drop in the wake behind the main cylinder, as shown in Figure~\ref{fig:CpCfCompared} (a), which in turn is responsible for the observed reduction in mean drag.
    The phenomenon observed also occurs in using the technique of boat tailing~\cite{reubush1976effect,lunghi2024drag}, a method widely recognized for its effectiveness in reducing drag on bluff bodies.
    However, the current approach achieves drag reduction by optimizing the positions of two smaller cylinders downstream of the main cylinder. 
    Adjusting their placements can form a large, stable separation vortex, which acts like a streamlined fairing behind the cylinder to effectively reduce drag. 
    It is in contrast to traditional boat tailing that relies on the geometric reshaping of bluff bodies.
    Compared to the uncontrolled case, the skin friction coefficient $C_f$ on the main cylinder with flow control is slightly reduced as shown in Figure~\ref{fig:CpCfCompared} (b), but its magnitude is relatively small, contributing approximately $6\%$ to the total drag coefficient. 
    Regarding drag vibrations, these are the consequence of the periodic shedding of Kármán vortices. 
    The small cylinders serve to disrupt the formation of large vortices before they fully develop, thereby preventing the emergence of shedding vortices and mitigating drag vibrations.

\begin{figure}[!htb]
    \centering
    \centering
    \subfloat[Flow without control]{
    \includegraphics[width=0.486\textwidth]{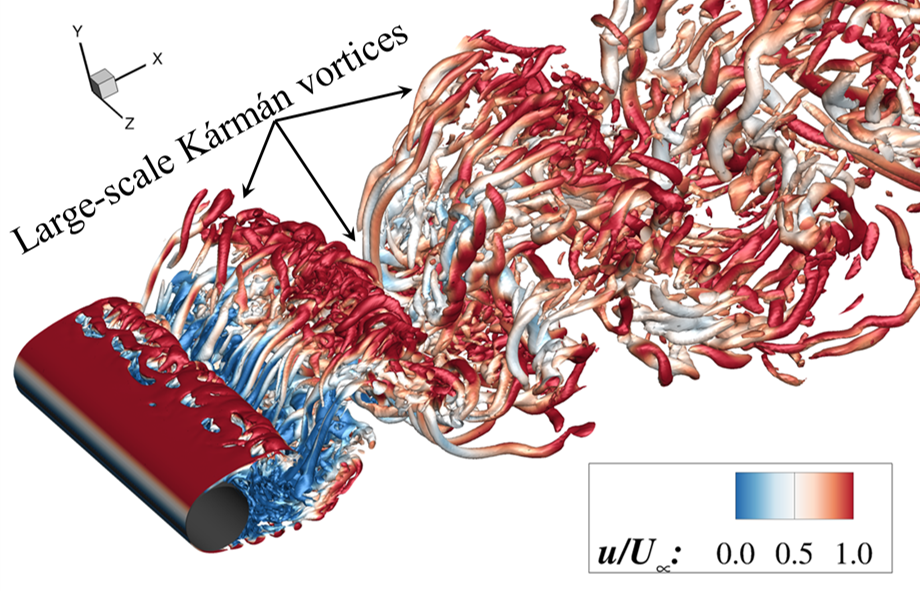} }
    \subfloat[Flow with control]{
    \includegraphics[width=0.486\textwidth]{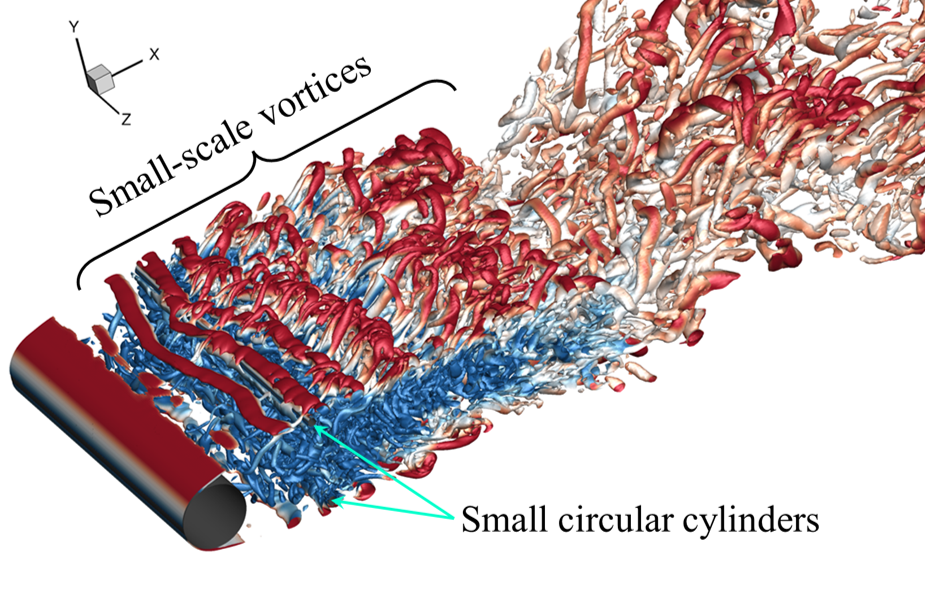} }
    \caption{ Visualization of vortical flows using $Q$-criterion ($Q$) isosurface based on the LES results: (a) without and (b) with control. $Q$ is defined as $Q = 0.5\left({{\bf{\Omega \Omega  - SS}}}\right)$. }
    \label{fig:LESCompared}
\end{figure}

    To validate the effectiveness of the optimized flow control strategy, we conduct high-fidelity large eddy simulations (LES) of the flow past a cylinder with two smaller downstream cylinders at $Re_D=3900$. 
    Figure~\ref{fig:LESCompared} presents the visualization of the vortical flow by Q-criterion isosurfaces in the wake, and the isosurfaces are colored by local $x$-axis velocity. 
    In Figure~\ref{fig:LESCompared} (a), we observe that the shear layers separate from the surface and encounter instability, leading to a sequence of Kármán vortices that increase in size as moving downstream.
    When the flow is controlled using two smaller cylinders, as shown in Figure~\ref{fig:LESCompared} (b), the shear layers originating from the main cylinder interact with these small cylinders. 
    As a result, numerous small vortices are formed instead of the large-scale Kármán vortices.
    The formation of these smaller vortices leads to a relatively streamlined and narrower wake, thereby reducing the pressure drag.
    The LES results demonstrate that the flow control with the small cylinders can achieve a $23.522\%$ reduction in $\langle{C_d}\rangle _\text{mean}$ and a $46.153\%$ reduction in $\langle{C_d}\rangle _{\text{std}}$, indicating the effectiveness of the optimized flow control with the ensemble Kalman method.
    
\subsection{Buffeting flow around NACA 0012 airfoil}
    \label{sec:3.2-NACA0012} 

    Flow over the NACA 0012 airfoil is one of the typical external transonic flows, which has been widely used for investigating transonic buffeting.
    The buffeting is characterized by a large-scale, self-sustained, low-frequency oscillation of the shock wave across the airfoil surface, resulting in significant vibrations in lift and drag forces, accompanied by intense pressure pulsations within the flow field~\cite{giannelis2017review}.
    This phenomenon occurs within specific ranges of the angle of attack ($\alpha$) and Mach number ($Ma$).
    For the NACA 0012 airfoil, intensive shock buffeting is reported at $\alpha=5.5^\circ$ and $Ma=0.7$ when the Reynolds number based on the freestream velocity and chord length is $3.0\times 10^6$ \cite{doerffer2010unsteady}. 
    This flow condition is used here to assess the effectiveness of the ensemble-based flow control in mitigating shock oscillations with a compliant aileron.

\begin{figure}[!htb]
    \centering
    \centering
    \subfloat[Numerical setup]{
    \includegraphics[height=0.355\textwidth]{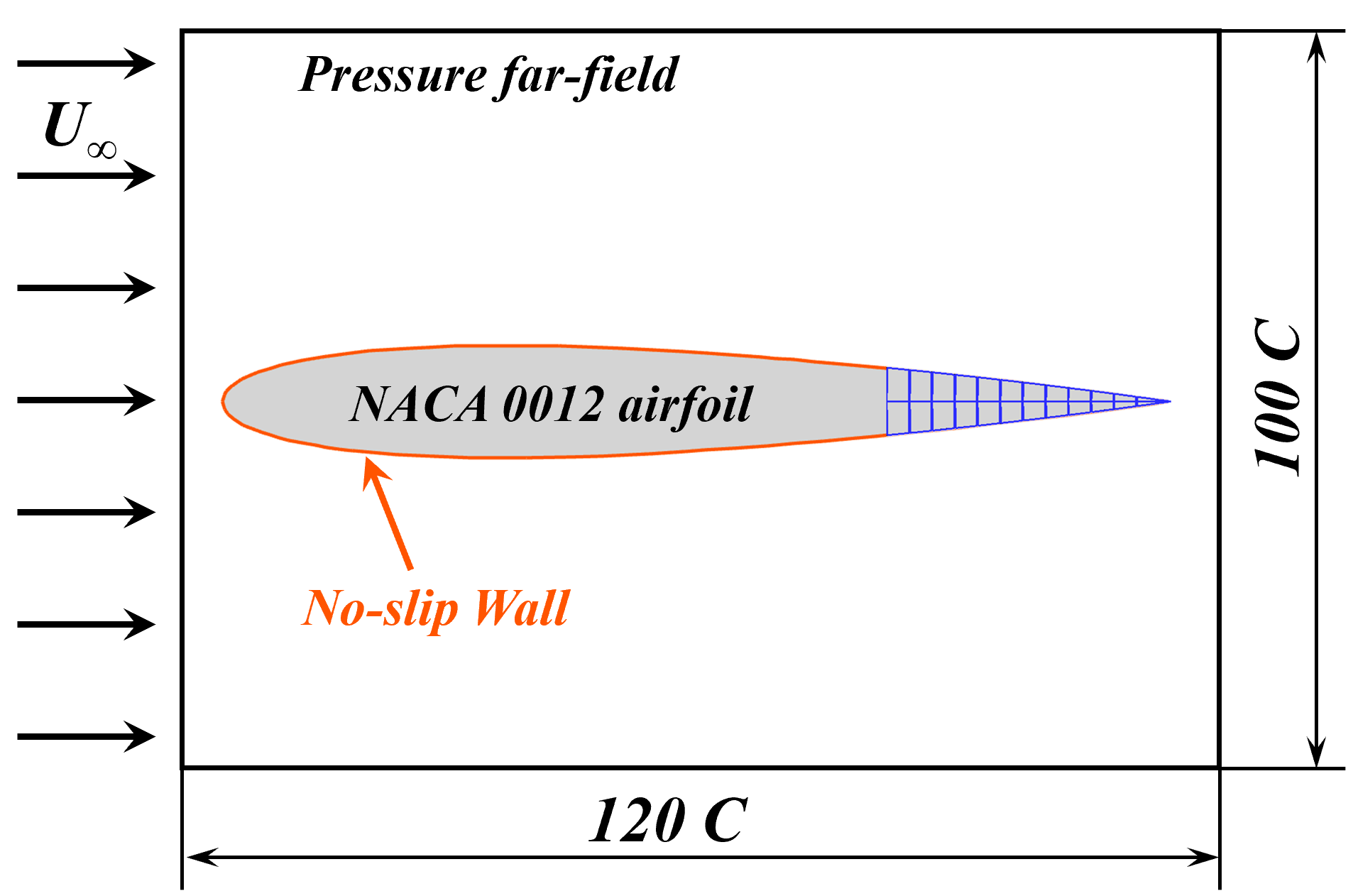} }
    \hspace{5mm}
    \subfloat[Computational mesh]{
    \includegraphics[height=0.335\textwidth]{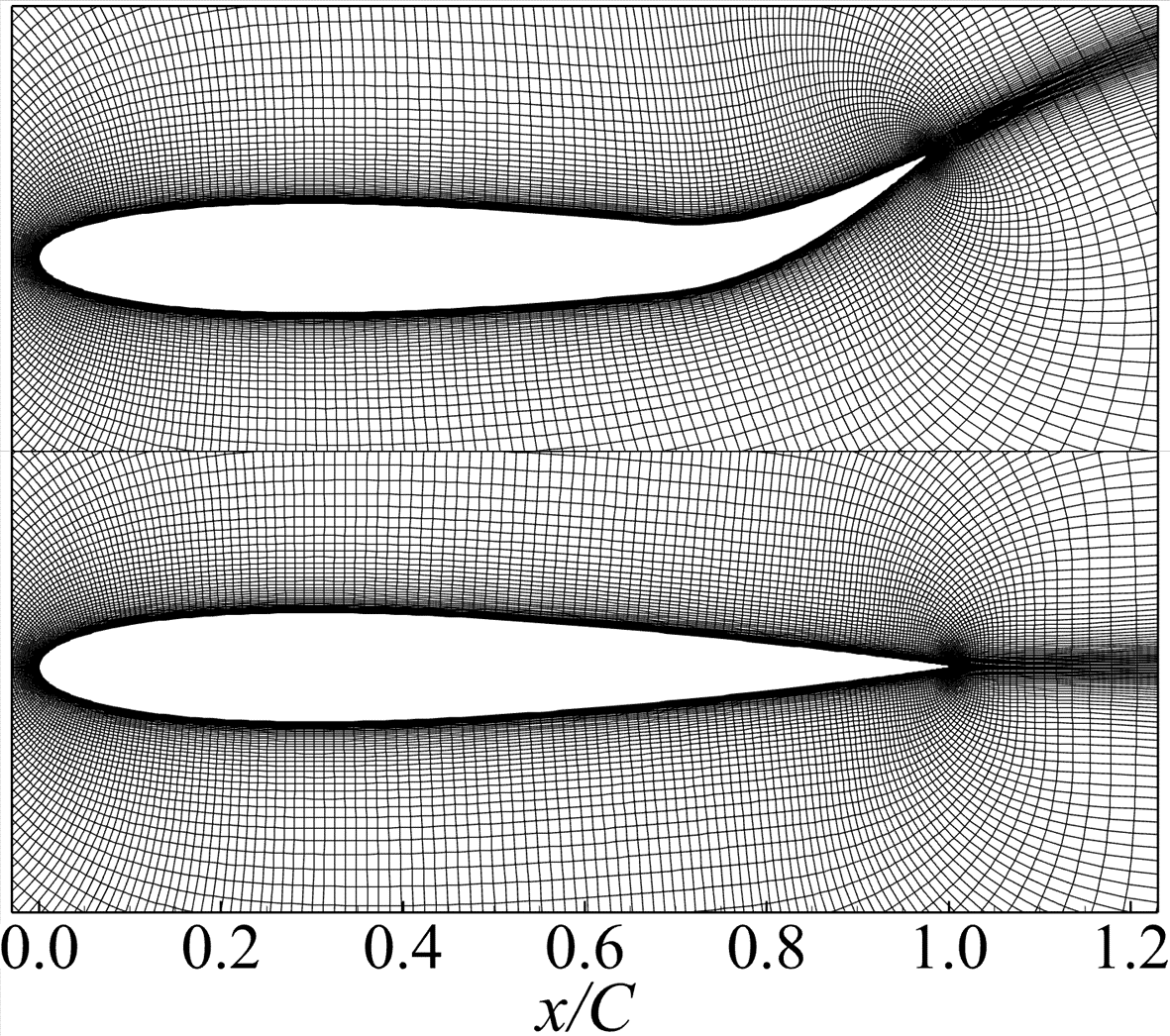} }
    \caption{ (a) The computational domain for the NACA 0012 airfoil. (b) Zoomed view of the mesh near the airfoil. }
    \label{fig:NACA0012Mesh}
\end{figure}

    Figure~\ref{fig:NACA0012Mesh}(a) shows the computational domain for the NACA 0012 airfoil, which is defined as a rectangular box with an extent of $x \in \left[ {-50C,70C} \right]$ and $y \in \left[ {-50C,50C} \right]$. 
    The leading edge of the airfoil is positioned at the origin.
    The adiabatic no-slip wall condition is applied on the airfoil surface, and the far-field is imposed with a non-reflective boundary condition.
    A grid convergence study has been conducted concerning the frequency of shock buffeting as shown in~\ref{sec:AppendixA}.
    The mesh deformation for the flapping motion of the trailing edge is achieved using a radial basis function (RBF) based interpolation method~\cite{wang2015improved}. 
    The corresponding governing equations on the moving mesh are solved using the arbitrary Lagrangian–Eulerian (ALE) method~\cite{blazek2015computational}. 
    In Figure~\ref{fig:NACA0012Mesh} (b), the top subfigure shows the trailing edge at a status of upward deflection, while the bottom one shows the trailing edge without deformation. 

\begin{figure}[!htb]
    \centering
    \subfloat[History of $C_l$, $C_d$, and $\Delta$]{
    \includegraphics[width=0.92\textwidth]{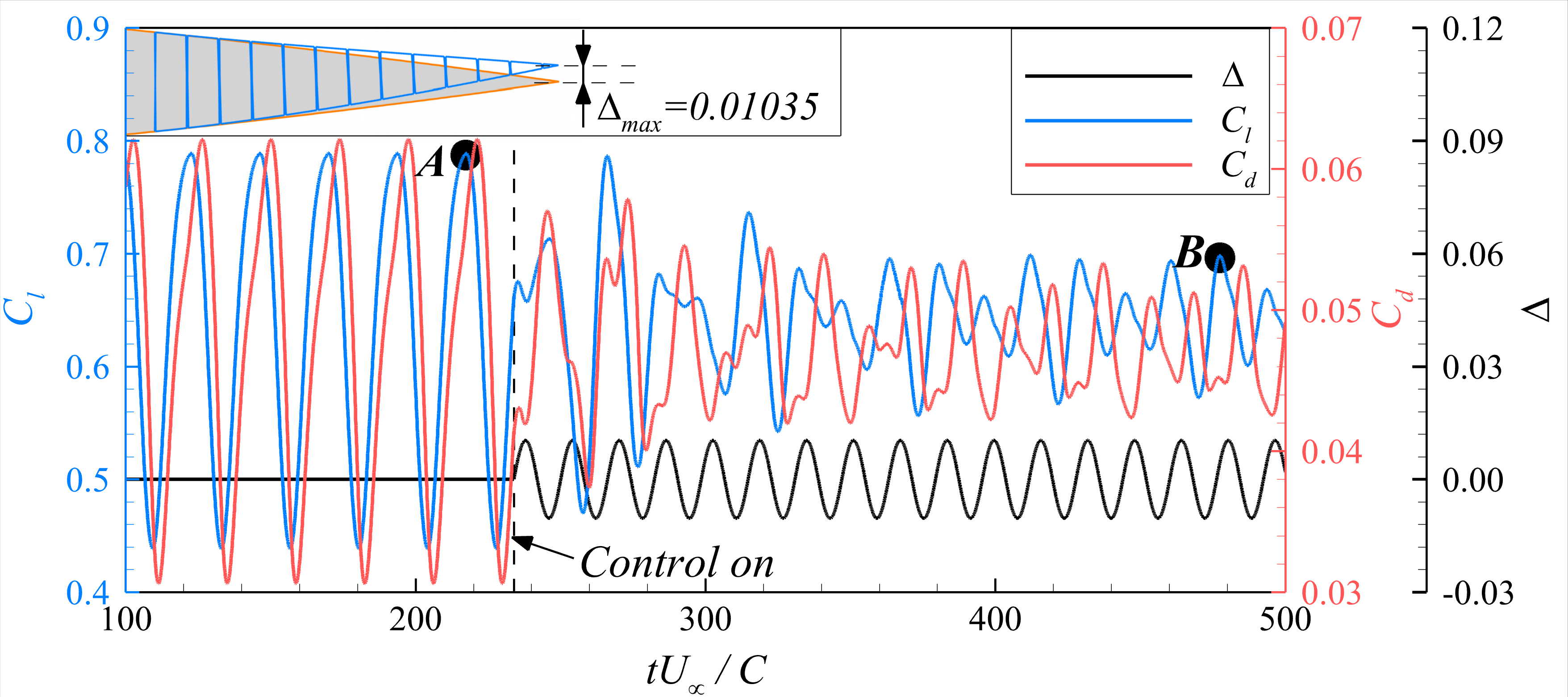} } \\
    \subfloat[Instantaneous $C_p$ contours at point A]{
    \includegraphics[width=0.465\textwidth]{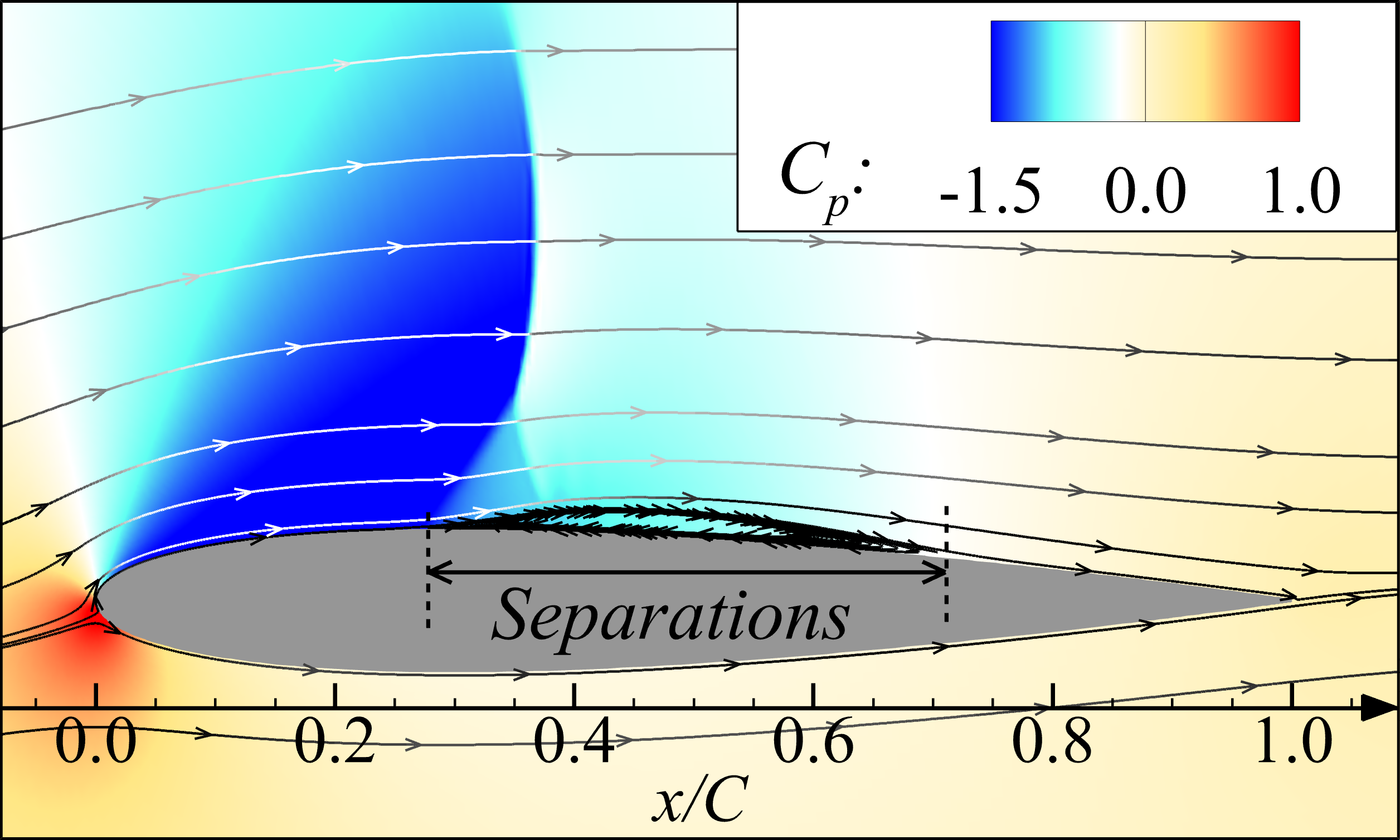} }
    \hspace{3mm}
    \subfloat[Instantaneous $C_p$ contours at point B]{
    \includegraphics[width=0.465\textwidth]{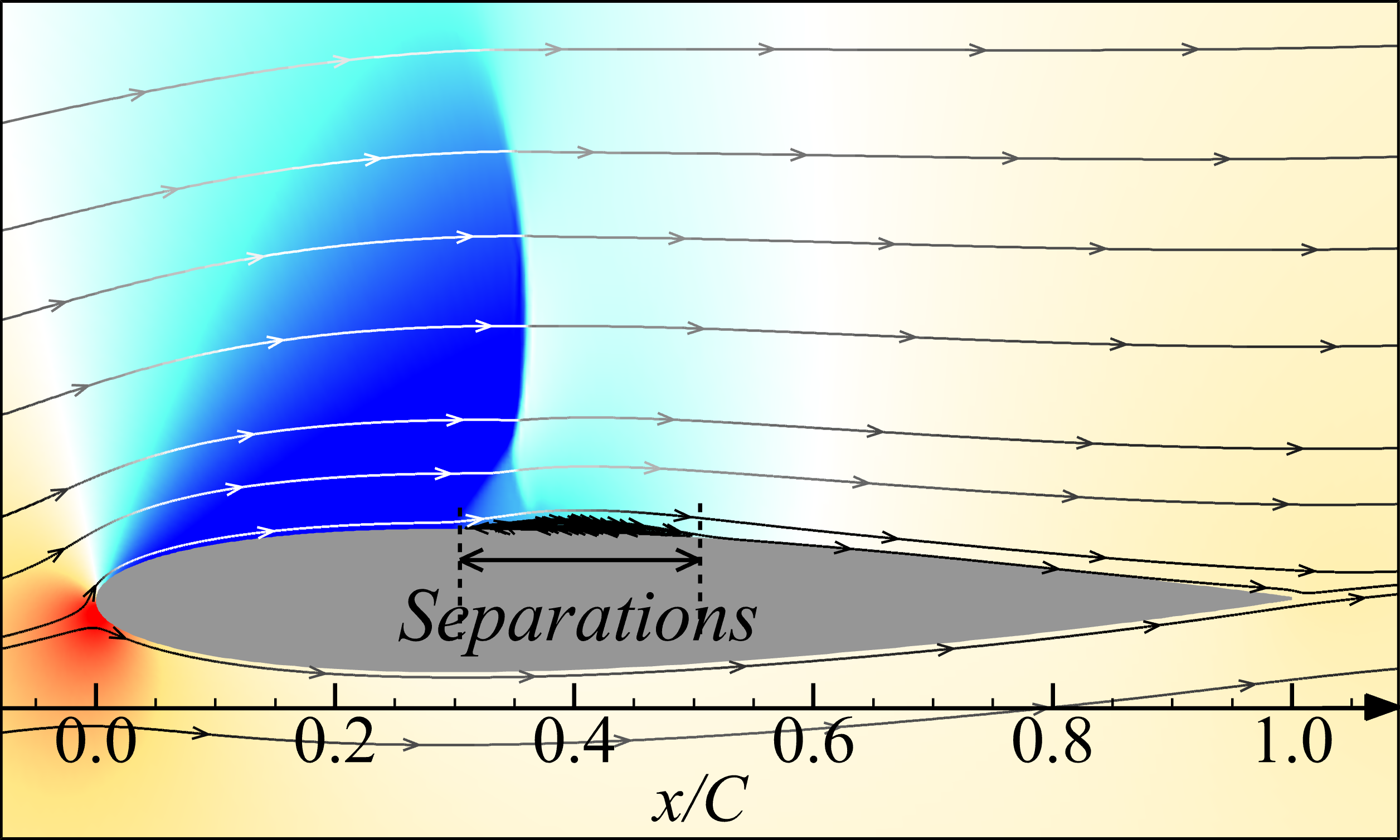} } 
    \caption{ (a) Time history of lift, drag, and displacement of trailing edge; (b, c) $C_p$--contours at maximum lift before (b) and after (c) the actuator activation. }
    \label{fig:NACA0012Cl}
\end{figure}

    The ensemble Kalman method is able to reduce the vibrations of both $C_l$ and $C_d$ by optimizing the movement of the trailing edge.
    The control parameters are optimized to be $\Delta_{max}=0.01035C, \eta=1.516, L=0.235C$, respectively. 
    Note that the optimal flapping frequency is approximately $1.5$ times the frequency of the shock oscillations in the uncontrolled flow case, which agrees with the observations in Ref.~\cite{caruana2003buffet}.
    With the actuator operating under these parameters, the vibration of $C_l$ is significantly reduced. 
    Figure~\ref{fig:NACA0012Cl} (a) shows the progressive decrease in the lift coefficient with the activation of the actuator.     
    The actuator is activated at the non-dimensional time $t^*=234$ ($t^*=t\cdot U_\infty/C$), marked with a black dashed line, initiating periodic flexible flapping of the trailing edge. 
    The maximum deformation (upperward deflection) of the trailing edge is indicated in the inset image. 
    The trailing edge displacement conforms to a sinusoidal function, as depicted by the black line. 
    After activating the actuator, the lift coefficient experiences damping of the flow-induced vibrations, ultimately stabilizing at a low-amplitude vibration. 
    A similar damping effect is observed for the drag coefficient. 
    Figure~\ref{fig:NACA0012Cl} (b) and (c) show contours of two typical instantaneous pressure coefficients $C_p$ along with streamlines.
    The two typical $C_p$--contours represent the instant of maximum lift coefficient before and after the activation of the actuator, corresponding to the points A and B in Figure~\ref{fig:NACA0012Cl} (a).
    With the actuator intervention, the shock is slightly shifted downstream, while the separation at the shock foot is considerably compressed from a span of $x/C \in [0.281,0.709]$ to a reduced span of $x/C \in [0.311,0.501]$.
    Consequently, the length of the separation bubble is shortened by $55.6\%$.

    Table~\ref{tab:ErrorEstmtNACA} summarizes the mean and standard deviations of both $C_l$ and $C_d$, as well as the reductions achieved through actuator intervention. 
    It is evident that the airfoil equipped with an actuator is superior to the uncontrolled case in terms of the vibration of both $C_l$ and $C_d$. 
    The standard deviations of lift coefficient ($\langle{C_l}\rangle _{\text{std}}$) and drag coefficient ($\langle{C_d}\rangle _{\text{std}}$) is reduced from  $1.153 \times 10^{-1}$ and $9.113 \times 10^{-3}$ to $3.108 \times 10^{-2}$ and $2.935 \times 10^{-3}$, resulting in reductions of $73.040\%$ and $67.786\%$, respectively.
    The mean lift coefficients ($\langle{C_l}\rangle _\text{mean}$) and drag coefficients ($\langle{C_d}\rangle _\text{mean}$) shows marginal changes, with an increase of $0.664\%$ and a decrease of $1.247\%$, respectively.

\begin{table}[!htb]
    \caption{ \label{tab:ErrorEstmtNACA} 
    Summary of changes in mean and standard deviation of lift and drag coefficients. }
    \centering
    \begin{tabular}{cccccc}
    \hline
    \hline
                 & \multirow{2}{*}{Uncontrolled}   & \multicolumn{4}{c}{Active controlled} \\
    \cline{3-6}
                                  &                      & Open-loop      & Relative change ($\epsilon_\text{rel}$)   & Closed-loop     & Relative change ($\epsilon_\text{rel}$) $^a$ \\
    \hline
    $\langle{C_l}\rangle_\text{mean}$  & $6.389\times10^{-1}$ & $6.431\times10^{-1}$ & $0.664\%  $ & $6.249\times10^{-1}$ & $-2.191\%  $  \\
    $\langle{C_l}\rangle_{\text{std}}$  & $1.153\times10^{-1}$ & $3.108\times10^{-2}$ & $-73.040\%$ & $8.656\times10^{-4}$ & $-99.249\% $  \\
    $\langle{C_d}\rangle_\text{mean}$  & $4.779\times10^{-2}$ & $4.719\times10^{-2}$ & $-1.247\% $ & $4.612\times10^{-2}$ & $-3.494\%  $  \\
    $\langle{C_d}\rangle_{\text{std}}$  & $9.113\times10^{-3}$ & $2.935\times10^{-3}$ & $-67.786\%$ & $1.601\times10^{-3}$ & $-82.432\% $  \\
    \hline
    \hline
    \multicolumn{6}{l}{ $^a$ { The relative change ($\epsilon_\text{rel}$) is estimated based on Eq.~\ref{eq:Rerror}, and negative sign indicates a reduction.} }
\end{tabular}
\end{table}

\begin{figure}[!htb]
    \centering
    \centering
    \subfloat[$C_{p, \text{rms}}$, without control]{
    \includegraphics[width=0.465\textwidth]{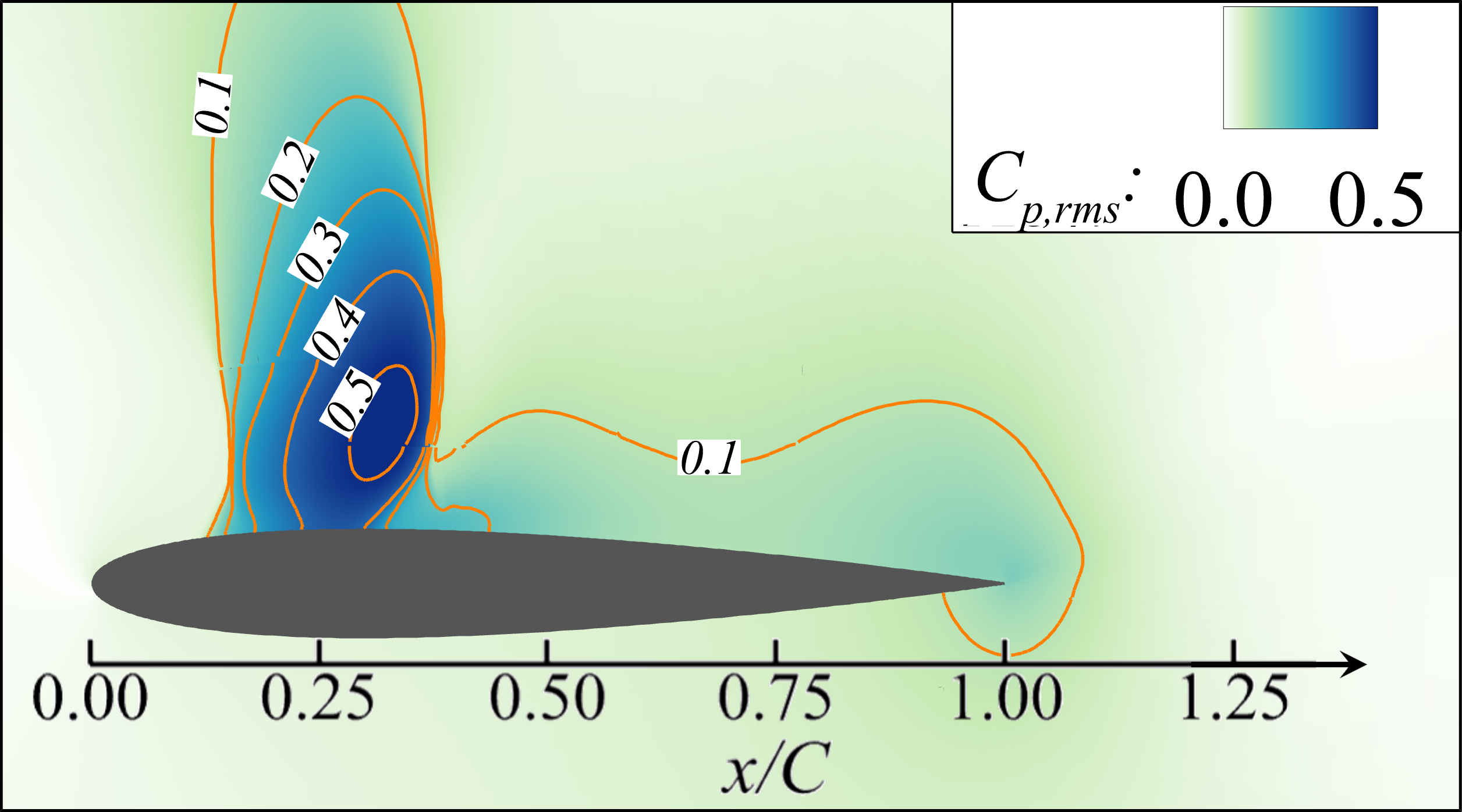} }
    \hspace{3mm}
    \subfloat[$C_{p, \text{rms}}$, with control]{
    \includegraphics[width=0.465\textwidth]{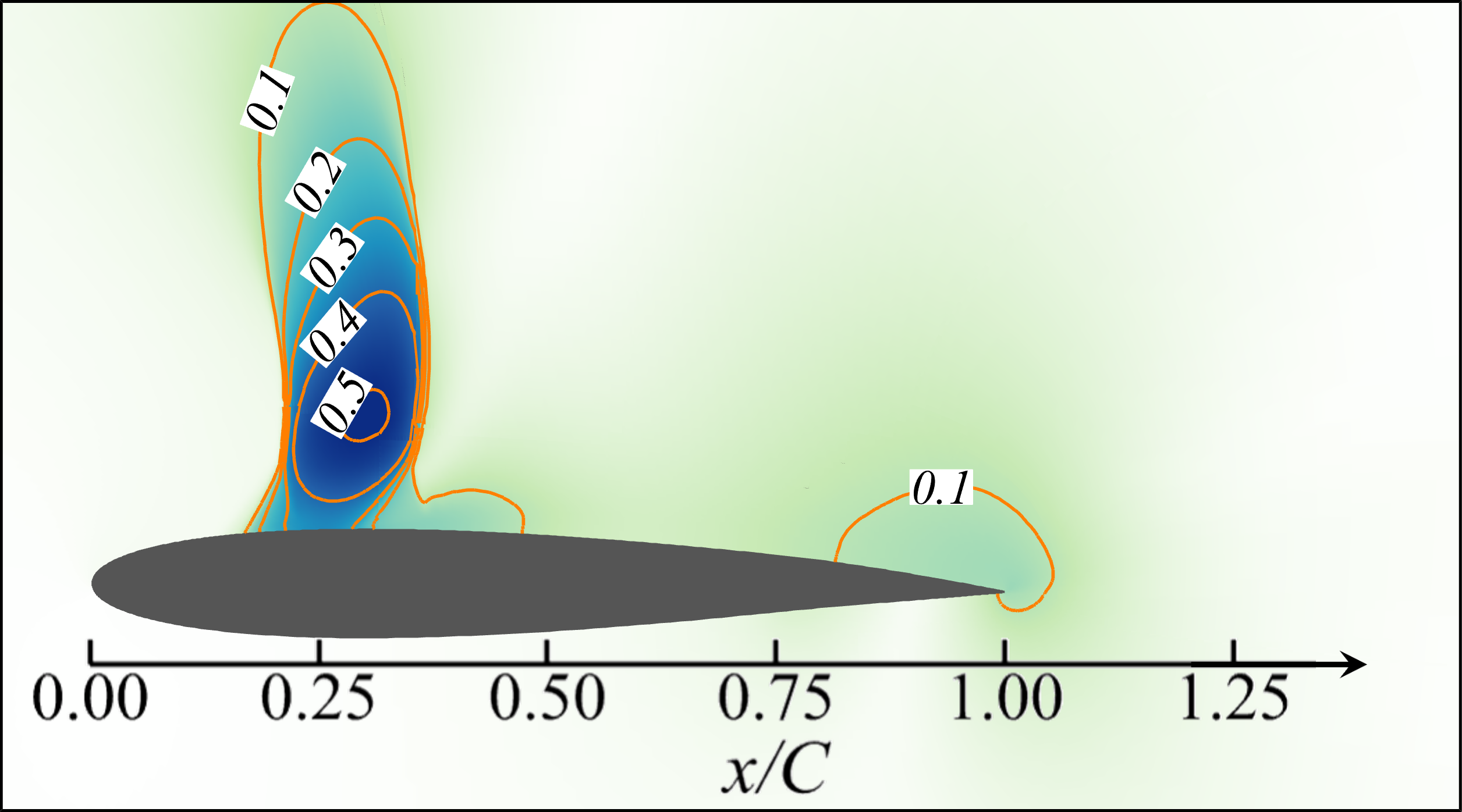} }
    \\
    \subfloat[$u_{\text{rms}}$, without control]{
    \includegraphics[width=0.465\textwidth]{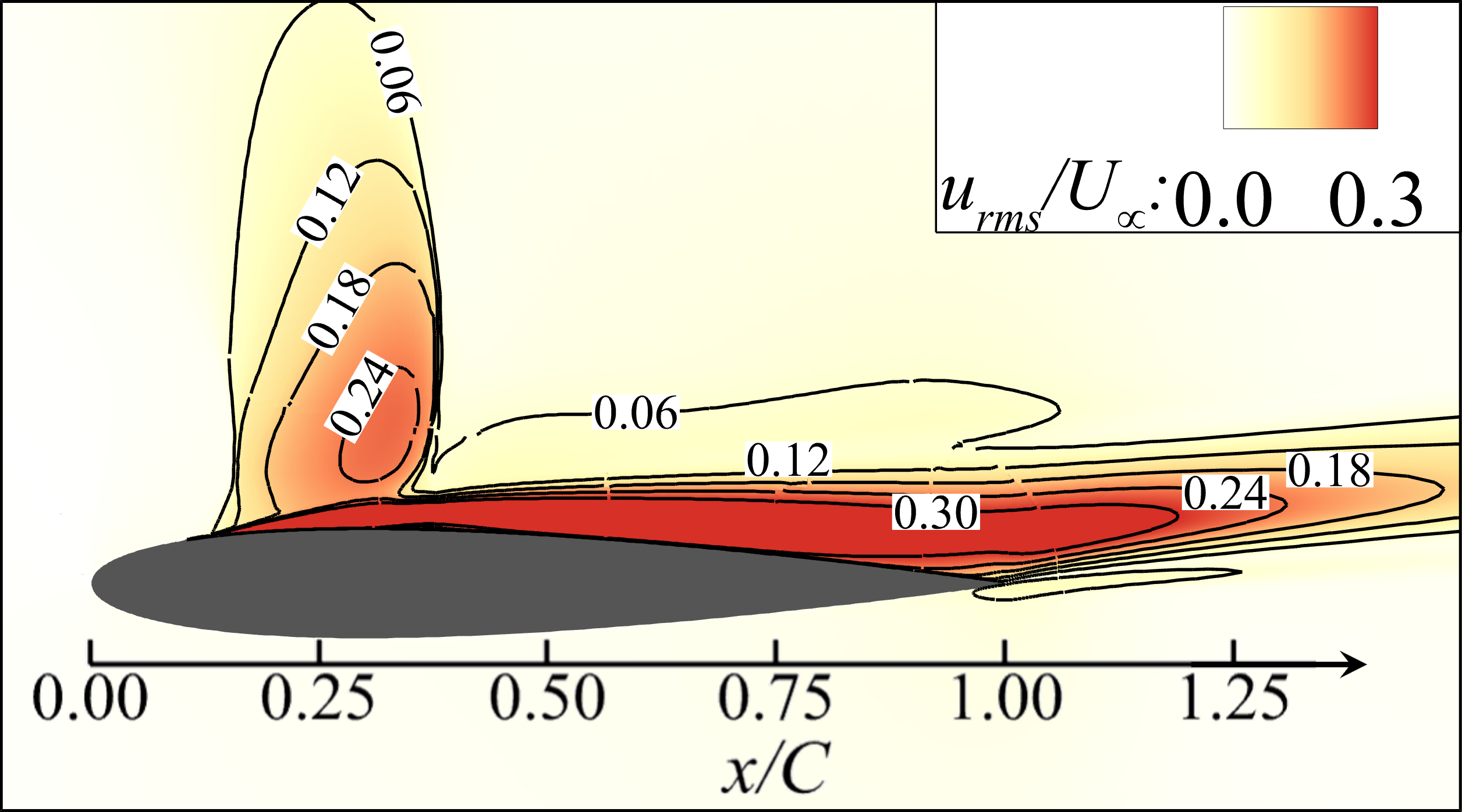} }
    \hspace{3mm}
    \subfloat[$u_{\text{rms}}$, with control]{
    \includegraphics[width=0.465\textwidth]{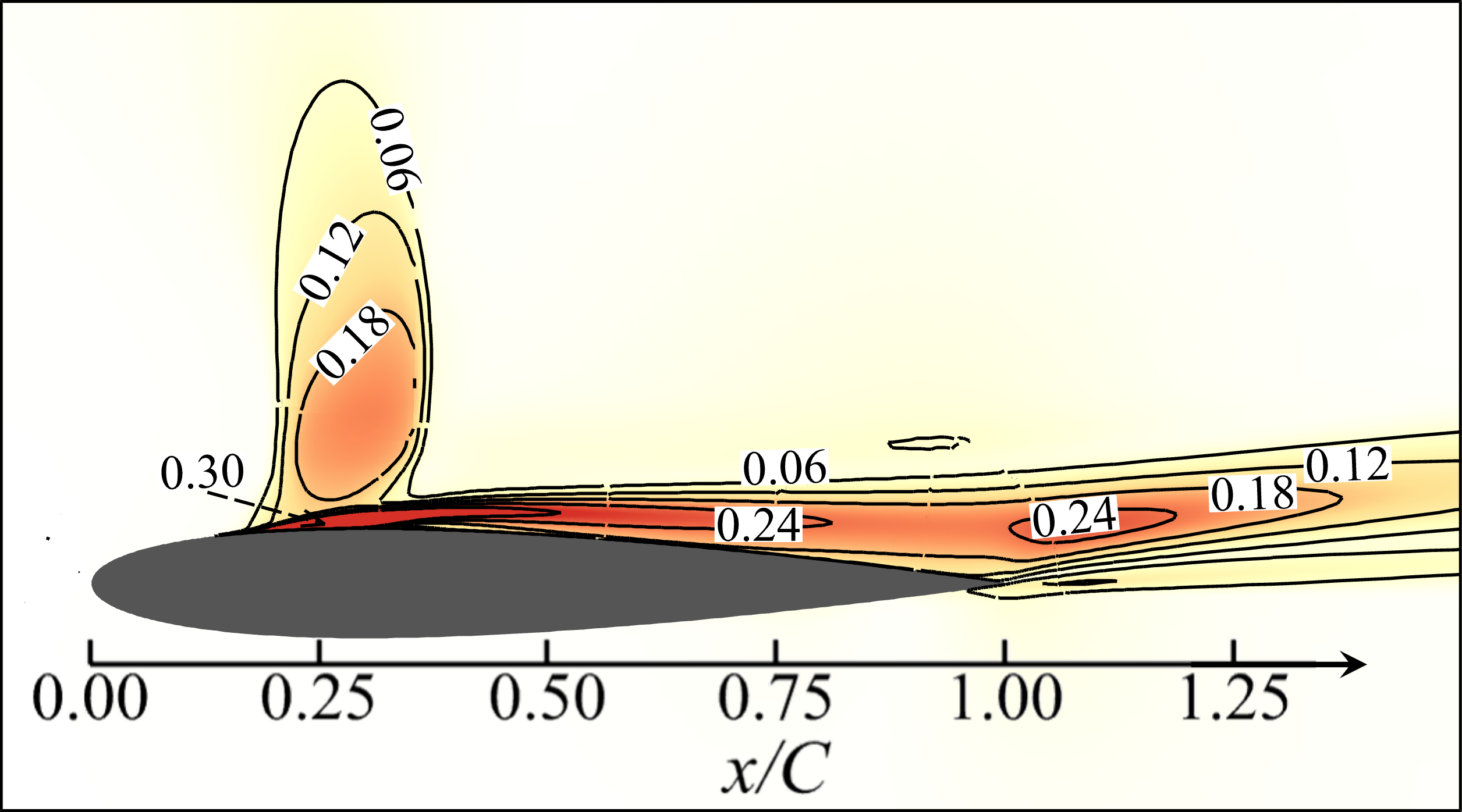} }
    \caption{ Contours of the root mean square (RMS) pressure ($C_{p,\text{rms}}$) and velocity ($u_{\text{rms}}$) for the NACA 0012 airfoil: (a, c) without control; (b, d) with open-loop control. }
    \label{fig:NACA0012Cprms}
\end{figure}

    The optimized flow control with the ensemble-based method can noticeably suppress the oscillation of shock waves. 
    Figure~\ref{fig:NACA0012Cprms} (a) and (b) illustrate the distributions of the RMS of the pressure coefficient ($C_{p,\text{rms}}$).
    On the upper surface, a bar-shaped region exhibits a relatively high level of $C_{p,\text{rms}}$ with $C_{p,\text{rms}} \geq 0.1$, induced by the shock oscillation. 
    In the case of flow control, the range of the shock oscillation is compressed from $x/C \in [0.132,0.388]$ to $x/C \in [0.197,0.367]$. 
    A similar scenario is depicted by the distributions of the RMS of the longitudinal velocity ($u_{\text{rms}}$) in Figure~\ref{fig:NACA0012Cprms} (c) and (d). 
    Furthermore, the uncontrolled case, as shown in Figure~\ref{fig:NACA0012Cprms} (c), indicates that the highest velocity fluctuations occur in the interaction region between the shock wave and the boundary layer and in the vicinity of the wake.
    Compared to the uncontrolled case, a significant decrease in the velocity fluctuation can be observed in the shock area external to the boundary layer in Figure~\ref{fig:NACA0012Cprms} (d).
    Notably, within the boundary layer, the region of $u_{\text{rms}} \geq 0.3$ is significantly compressed from the span of $x/C \in [0.15, 1.19]$ to $x/C \in [0.18, 0.51]$, resulting in an approximate $68.3\%$ reduction.

\begin{figure}[!htb]
    \centering
    \includegraphics[width=0.93\textwidth]{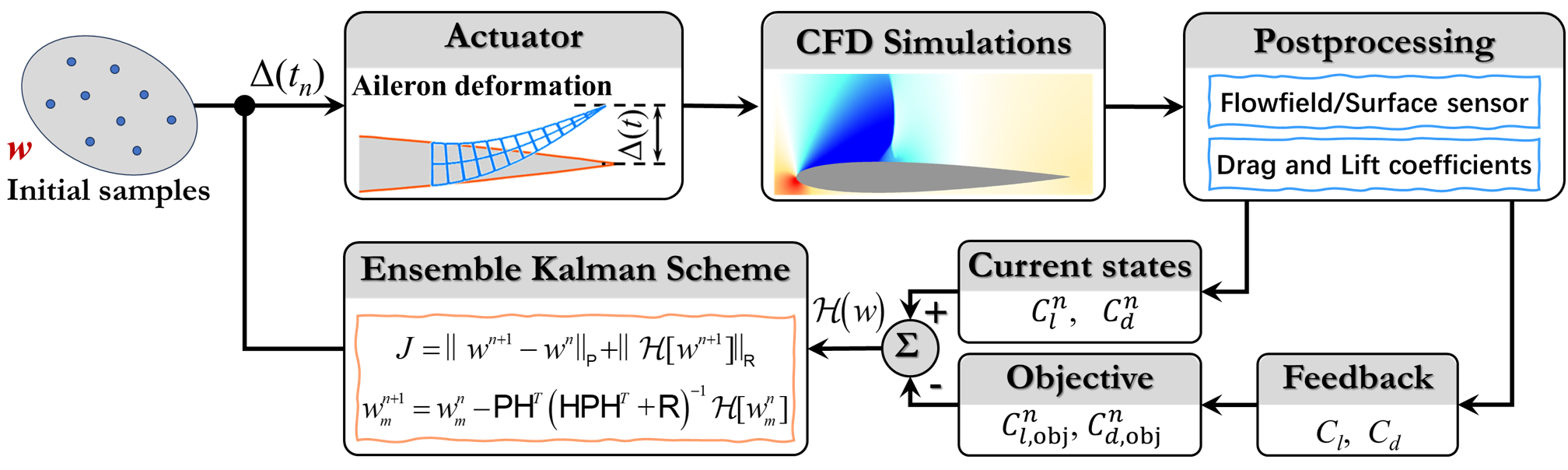} 
    \caption{Block diagram for the closed-loop flow control.} 
    \label{fig:schematicClosedLoop}
\end{figure}

    The ensemble-based method is also tested to optimize the closed-loop active control strategy for the transonic flow over the NACA 0012 airfoil.
    The length of the aileron is fixed at $0.235C$ in this case, which is the same as that used in the open-loop control application. 
    In the closed-loop control, the aileron movement is not confined to any specific functional form.
    The block diagram of the closed-loop control is shown in Figure~\ref{fig:schematicClosedLoop}.
    The actuator takes actions (deformation of ailerons) based on the real-time inputs $\Delta(t_n)$, where the subscript $n$ represents the index of the time step. 
    The impact of aileron deformation on the flow is analyzed using CFD simulations, which provide the response of flow fields.
    The historical lift and drag signals~$C_l(t_{n-1})$ and $C_d(t_{n-1})$ prior to the current step $n$ are used as feedback signals for estimating the objectives of lift and drag coefficients, i.e., $C_{l,\text{obj}}$ and $C_{d,\text{obj}}$, for the next loop.
    These simulations undergo postprocessing to extract the current states, i.e., the $C_l$ and $C_d$ at the current time step.
    The current states are used to construct the observed quantities $\mathcal{H}[w]$ by analyzing their differences with the objectives~$C_{l,\text{obj}}$ and $C_{d,\text{obj}}$.
    The observed quantities are then fed into the Kalman update scheme to update the aileron deformation $\Delta(t_n)$.   
    Subsequently, aileron deflection is driven by the ensemble-based method to control the actuator, steering the current lift and drag coefficients toward the objective values in the next loop. 
    By incorporating the feedback signals, the system forms a closed-loop control mechanism, maintaining optimal performance through continuous adjustments.
    
    The objective of lift and drag coefficients at $n$--th time steps are represented by $C_{l,\text{obj}}^n$ and $C_{d,\text{obj}}^n$ respectively, which are calculated using the following equations:
\begin{equation}
    { \left\{ \begin{array}{l} 
    C_{l,\text{obj}}^n = \frac{1}{{n - {n_s} + N}}\left( {\sum\limits_{i = {n_s} - N}^{{n_s}} {C_l^i}  + \sum\limits_{i = {n_s} + 1}^{n - 1} {C_l^i} } \right) \\
    C_{d,\text{obj}}^n = \frac{1}{{n - {n_s} + N}}\left( {\sum\limits_{i = {n_s} - N}^{{n_s}} {C_d^i}  + \sum\limits_{i = {n_s} + 1}^{n - 1} {C_d^i} } \right)
    \end{array} \right. } \text{,}
\label{eq:ClOpt}
\end{equation}
    where ${n_s}$ is the time step for activation of the actuator, $N$ is the number of time steps before the actuator activation, $C_l^i$ and $C_d^i$ represent the lift and drag coefficients at $i$--th time steps.      
    In other words, we use the time-averaged lift and drag from the $N$ time steps before control and $n-n_s$ time steps after control, as the objective. 
    Further, the observed quantities $\mathcal{H}[w]$ are defined as the differences between the current states and the constructed objectives. 
    Specifically, ${\mathcal H}\left[ {{w^{n + 1}}} \right]$ is expressed as ${\mathcal H}\left[ {{w^{n + 1}}} \right] = {\left[ {\left( {C_l^n - C_{l,\text{obj}}^n} \right),\left( {C_d^n - C_{d,\text{obj}}^n} \right)} \right]^\top}$.
    In doing so, the objective value is dynamically adjusted at each time step to measure possible minimal flow vibration during the real-time control process.

\begin{figure}[!htb]
    \centering
    \includegraphics[width=0.93\textwidth]{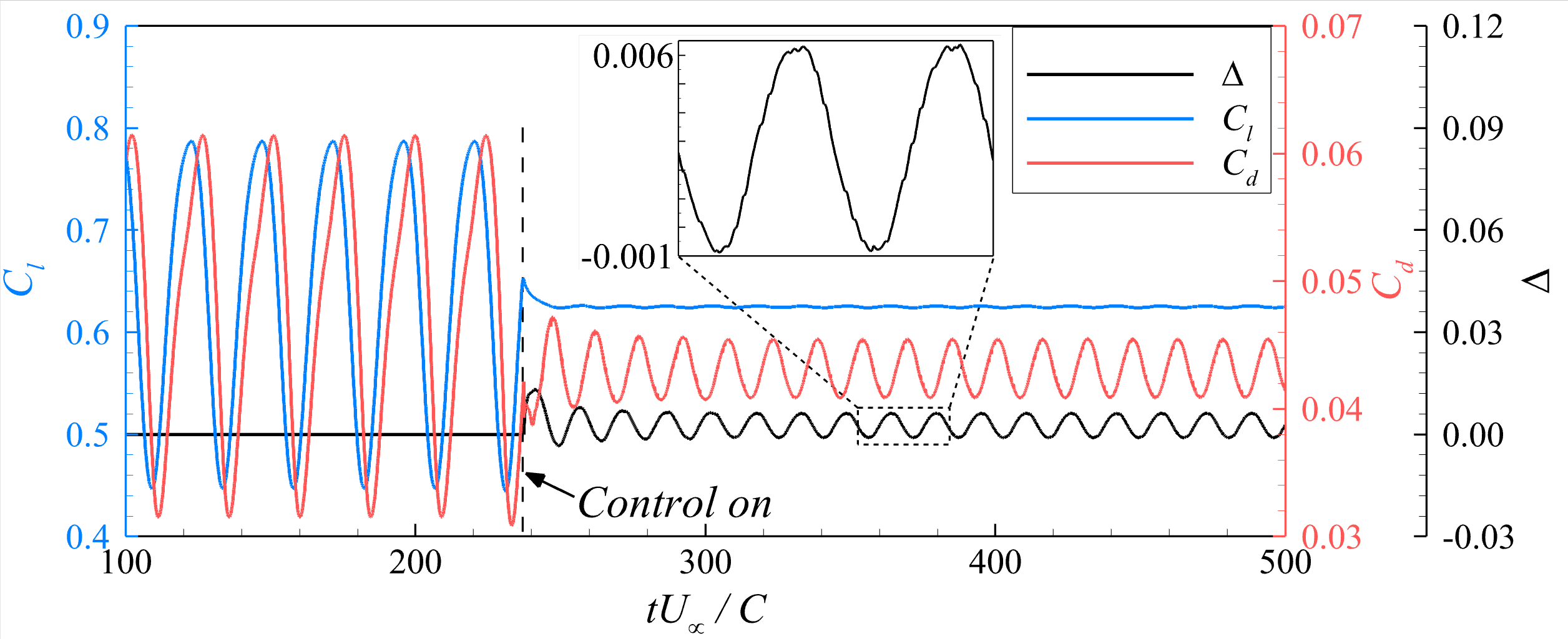}
    \caption{ Time history of $C_l$, $C_d$, and the displacement of trailing edge ($\Delta$), with the black dotted line indicating the closed-loop control activated. }
    \label{fig:NACA0012ClAdaptContl}
\end{figure}

    The closed-loop control optimized with the ensemble method demonstrates a significant enhancement in vibration mitigation compared to the open-loop control.
    Figure~\ref{fig:NACA0012ClAdaptContl} illustrates the time history of the lift coefficient ($C_l$), the drag coefficient ($C_d$), and the displacement of the trailing edge ($\Delta$). 
    The activation of the closed-loop control is indicated by the black dotted line. 
    Once activated, the amplitude of both lift and drag coefficients decays immediately. 
    The lift $C_l$ exhibits little variation over time, and its standard deviation is reduced from $1.153 \times 10^{-1}$ to $8.656 \times 10^{-4}$, resulting in a reduction of $99.249\%$, as listed in Table~\ref{tab:ErrorEstmtNACA}. 
    The $C_d$ exhibits reduced periodic vibrations, which is an expected consequence of the ailerons suppressing the shock motion through flapping.
    The corresponding standard deviation of $C_d$ is decreased by $82.432\%$ as listed in Table~\ref{tab:ErrorEstmtNACA}.
    The displacement of the trailing edge ($\Delta$), indicated by black lines, is generally exhibited as a sinusoidal-like function, but it is superimposed with high-frequency signal components (as shown in the inset image). 
    The high-frequency signal comes mainly from the real-time adjustment of the actuator.
    The trailing edge displacement are ultimately stabilized within the range $\Delta/C\in[-9.5\times 10^{-4}, 6.2 \times 10^{-3}]$, corresponding to an amplitude of $7.15\times 10^{-3}C$.
    This represents a significant reduction of $65.46\%$ compared to the amplitude with the open-loop active control that has an amplitude of $2.07 \times 10^{-2}C$.

\begin{figure}[!htb]
    \centering
    \centering
    \subfloat[$C_{p,\text{rms}}$ contours]{
    \includegraphics[width=0.465\textwidth]{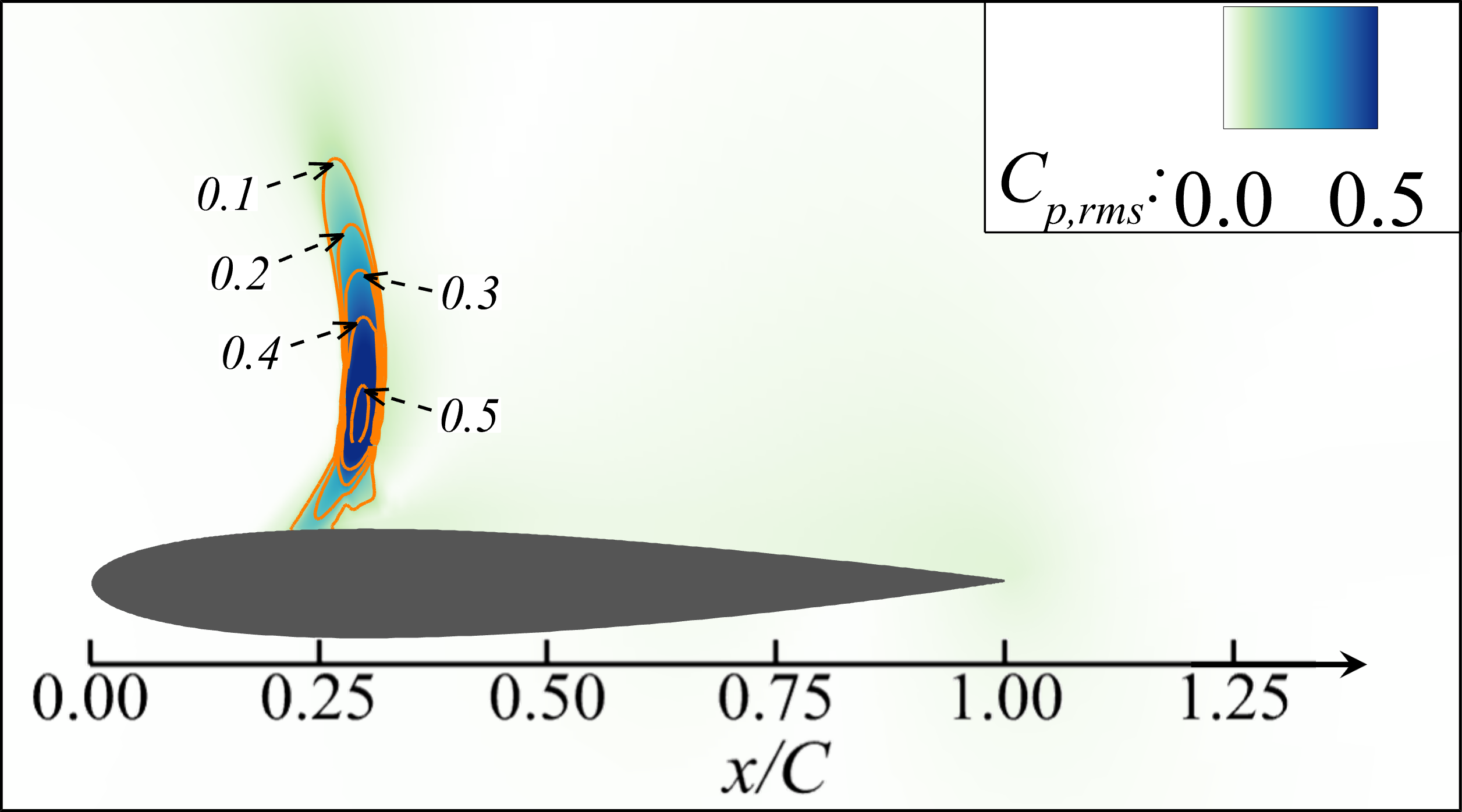} }
    \hspace{3mm}
    \subfloat[$u_{\text{rms}}$ contours]{
    \includegraphics[width=0.465\textwidth]{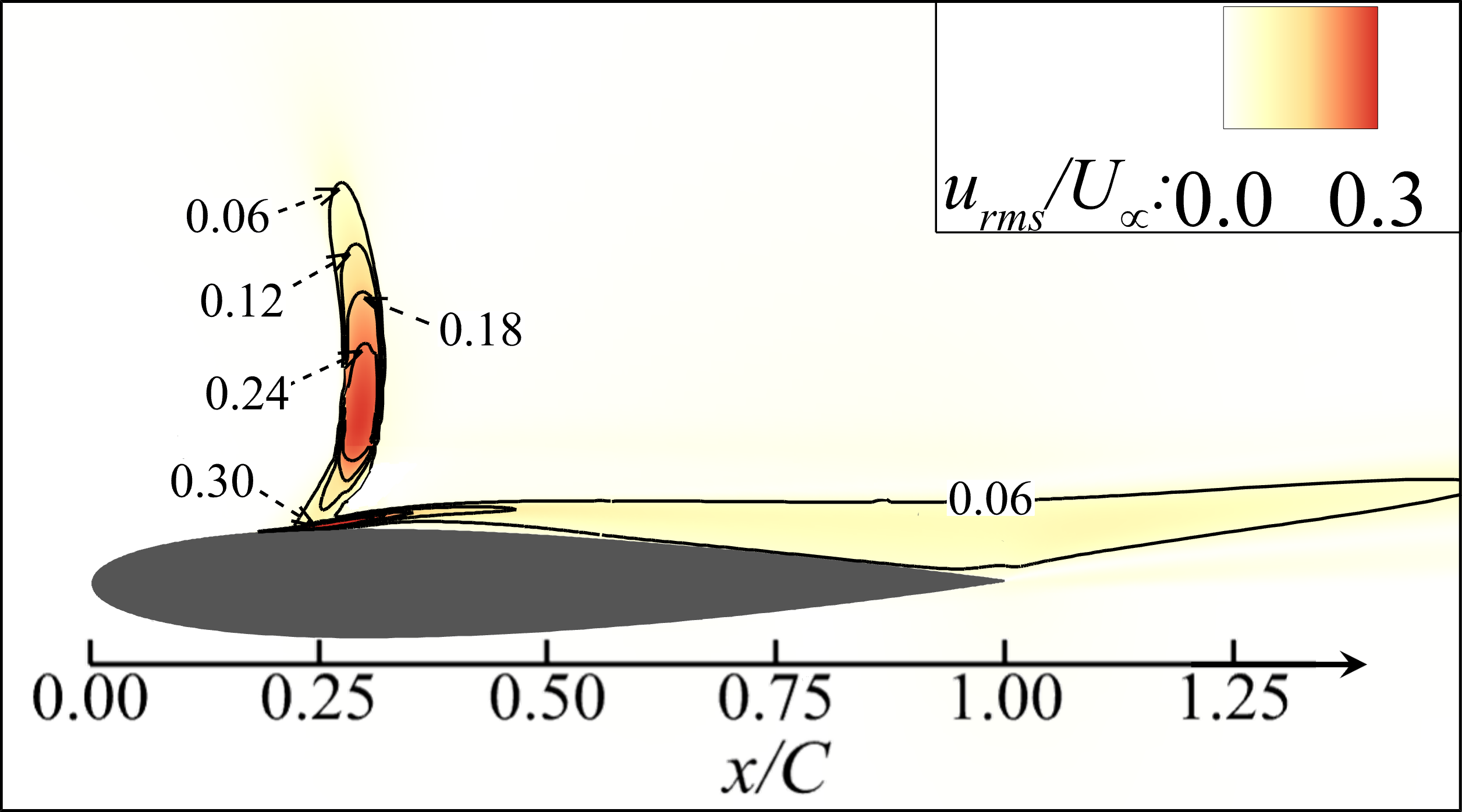} }
    \caption{ The RMS pressure ($C_{p,\text{rms}}$) (a) and velocity ($u_{\text{rms}}$) (b) contours around NACA 0012 with closed-loop control. }
    \label{fig:NACA0012CprmsUrmsAdaptContl}
\end{figure}

    The closed-loop control with the ensemble Kalman method is shown to be more effective at suppressing shock oscillations, in contrast to the open-loop control. 
    Figure~\ref{fig:NACA0012CprmsUrmsAdaptContl} presents contours of the RMS pressure ($C_{p,\text{rms}}$) (a) and velocity ($u_{\text{rms}}$) (b) around a NACA 0012 airfoil.
    The region with $C_{p,\text{rms}} \geq 0.1$ is likely induced by the shock excursion, which shrinks significantly to $x/C \in [0.277, 0.319]$ as shown in Figure~\ref{fig:NACA0012CprmsUrmsAdaptContl} (a). 
    The closed-loop active control achieves an $83.59\%$ decrease compared to the uncontrolled case of $x/C \in [0.132, 0.388]$ and a $75.29\%$ decrease compared to the open-loop active control of $x/C \in [0.197, 0.367]$.
    Similar to the distribution of $C_{p,\text{rms}}$, the region with the relatively high level of $u_{\text{rms}}$ is also shrunk as presented in Figure~\ref{fig:NACA0012CprmsUrmsAdaptContl} (b). 
    
\section{Conclusion}
\label{sec:4-Conclusion}
    In this work, the ensemble Kalman method is used to optimize actuator-based flow control for mitigating the flow-induced vibration of a circular cylinder and the transonic buffeting over the NACA 0012 airfoil.
    The method can effectively optimize the placement and movement of actuators in a non-intrusive way.
    This offers flexibility for handling different objectives in various flow control problems.
    The ensemble-based flow control optimization is tested in two cases: low-speed flow past a circular cylinder and the transonic buffeting flow over a NACA 0012 airfoil. 
    We show that the optimized passive control with the ensemble method effectively suppresses the vortex shedding downstream of the circular cylinder. 
    Further, in the test case of transonic buffeting flow over the NACA 0012 airfoil, the ensemble-based method is utilized to optimize both open-loop and closed-loop active control, highlighting its flexibility in different flow control applications. 
    The results show that the optimized flow control effectively mitigates shock oscillations, leading to a significant reduction in the vibrations of both lift and drag.

\appendix

\counterwithin{figure}{section}
\counterwithin{table}{section} %
\counterwithin{equation}{section} %
\renewcommand\thesection{Appendix \Alph{section}}
\renewcommand\thefigure{\Alph{section}\arabic{figure}}
\renewcommand\thetable{\Alph{section}\arabic{table}}
\renewcommand\theequation{\Alph{section}\arabic{equation}}

\section{Verification of the CFD code} \label{sec:AppendixA}

    The verification of the CFD code is performed for grid convergence utilizing the Grid Convergence Index (GCI) method~\cite{roache1994perspective}. 
    It is to verify that the equations are being solved correctly and that the solution is insensitive to the grid resolution.
    The GCI method is a standardized way to report grid convergence quality typically based on three meshes with varying resolutions, i.e., coarse, medium, and fine mesh grids.
    All simulations are performed under consistent flow conditions.

    For flows over a circular cylinder, we generate three meshes with different resolutions, each with twice the number of mesh cells as the previous mesh.
    The first layer spacing in the normal direction for all meshes is chosen $y^+ \approx 0.6$, and the growth rate in the boundary layer is 1.1. 
    All grids are locally refined in the vortex-shedding areas to properly model the shear layer motion.
    Those meshes are referred to as coarse, medium, and fine grids, which consist of $34695$, $65238$, and $137606$ mesh cells, respectively. 
    Table~\ref{tab:MeshCylinder} summarizes the details of mesh information. 
    The effects of the grid resolution on Strouhal shedding frequency~$St$ are also presented in the table.
    The Strouhal number ($St$) is defined as $St = \frac{f L_\text{ref}}{{U_\infty }}$,
    where $f$ is the frequency of vortex shedding, $U_{\infty}$ represents the fluid velocity at the far field, and $L_\text{ref}$ represents the reference length.

\begin{table}[!htb]
    \caption{The mesh information for flows over circular cylinders}
    \label{tab:MeshCylinder}
    \centering
    \begin{tabular}{p{2.5cm}p{2cm}p{2cm}p{2cm}}
    \hline
    \hline
                    & Coarse  & Medium &   Fine  \\
    \hline   
    Number of cells &  34,695 & 65,238 & 137,606 \\
     $St$$^a$       &  0.2305 & 0.2230 &  0.2192 \\
    \hline
    \hline
    \multicolumn{4}{l}{$^a$ the reference length $L_\text{ref}$ is specified as the diameter of the cylinder ($D$). } \\
    \end{tabular}    
\end{table} 

\begin{figure}[!htb]
    \centering
    \subfloat[Circular cylinder]{\includegraphics[height=0.37\textwidth]{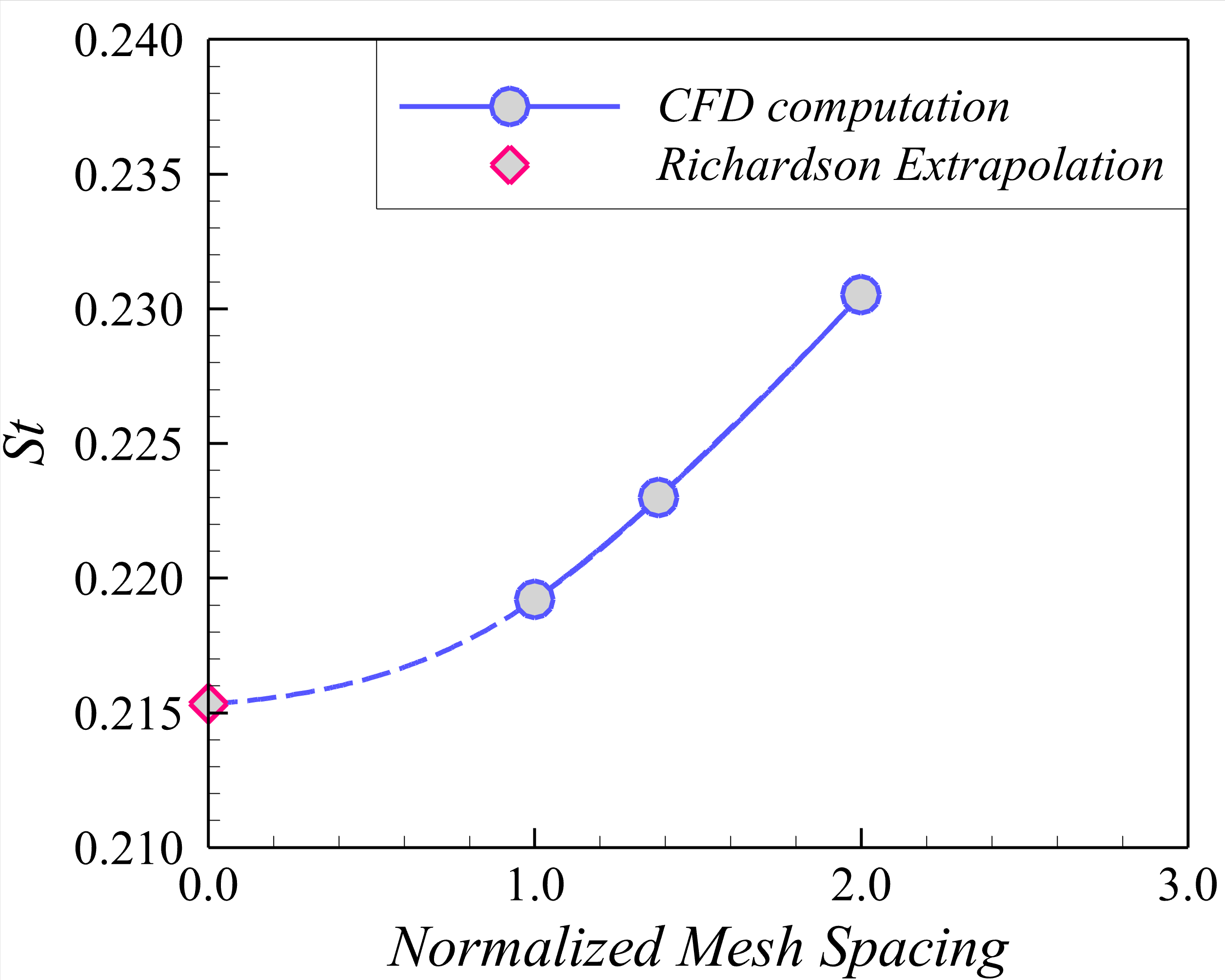}}
    \subfloat[NACA 0012 airfoil]{\includegraphics[height=0.37\textwidth]{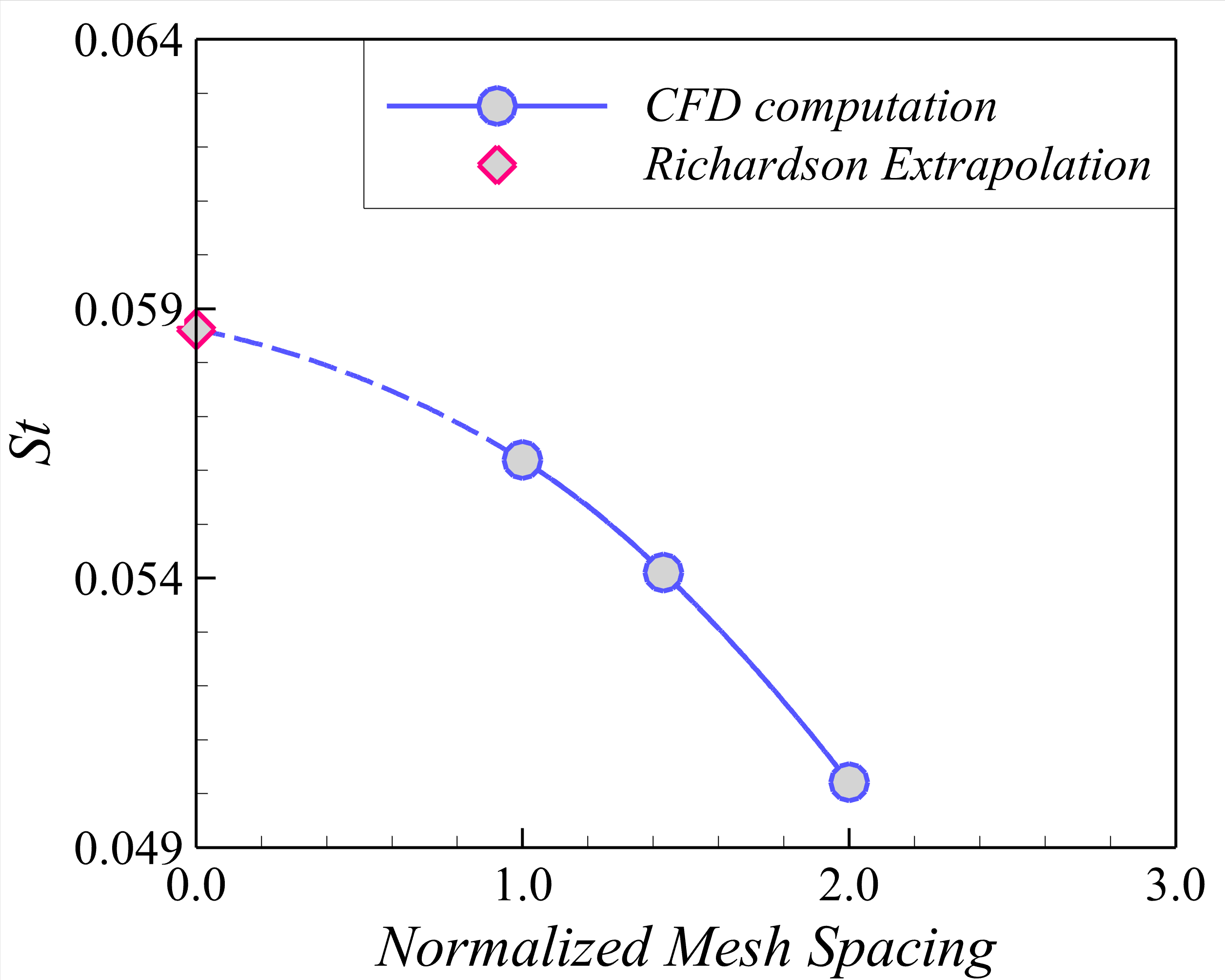} }
    \caption{Comparison of Strouhal number ($St$) with various mesh sizes for both circular cylinder (a) and NACA 0012 airfoil (b).}
    \label{figA:CylinderMeshStudy}
\end{figure}

    According to the GCI analysis, the order ($p$) of convergence based on the $St$ is estimated by
    \begin{equation}
        p = \frac{{\ln \left[ {\left( {S{t_{{\text{coarse}}}} - S{t_{{\text{medium}}}}} \right)/\left( {S{t_{{\text{medium}}}} - S{t_{{\text{fine}}}}} \right)} \right]}}{{\ln \left( {{r_{{\text{eff}}}}} \right)}} = 1.8326 \text{,}
    \label{eq:P-Oder}
    \end{equation}
    where $r_\text{eff}$ is the effective mesh refinement ratio and defined as ${r_\text{eff}} = {\left( {{{{N_\text{fine}}} \mathord{\left/ {\vphantom {{{N_\text{fine}}} {{N_\text{medium}}}}} \right. \kern-\nulldelimiterspace} {{N_\text{medium}}}}} \right)^{{1 \mathord{\left/  {\vphantom {1 D}} \right. \kern-\nulldelimiterspace} D}}} = 1.4523$. 
    $N$ is the total number of mesh cells. 
    Therefore, the theoretical convergence order is determined to be $p\approx 2$. 
    The value of $St$ at zero mesh spacing can be estimated by using the Richardson extrapolation: 
    \begin{equation}
        St_\text{asymptotic}={St_\text{fine}}+\frac{{{St_\text{fine}} - {St_\text{medium}}}}{{{r_\text{eff}^p} - 1}} = 0.2153 \text{.}
    \label{eq:R-extrapolation}
    \end{equation}
    Figure~\ref{figA:CylinderMeshStudy} (a) plots the $St_\text{asymptotic}$ and the predicted Strouhal numbers at different mesh spacing, where the $x$-axis is the spacing normalized by the spacing of the fine mesh. 
    It can be seen that the predictions of $St$ get close to $St_\text{asymptotic}=0.2153$ as the mesh spacing reduces.
    
    The mesh convergence index for both the fine mesh solution ($GCI_\text{f2m}$) and medium mesh solution ($GCI_\text{m2c}$) can be computed by: 
    \begin{equation}
       GCI_\text{f2m} = \frac{{{F_s}\left( {{St_\text{fine}} - {St_\text{medium}}} \right)}}{{{St_\text{fine}}\left( {{r_\text{eff}^p} - 1} \right)}} = 0.022076 \quad  
       GCI_\text{m2c} = \frac{{{F_s}\left( {{St_\text{medium}} - {St_\text{coarse}}} \right)}}{{{St_\text{medium}}\left( {{r_\text{eff}^p} - 1} \right)}} = 0.043001 \text{.}
    \label{eq:GCI}
    \end{equation}
    Here, the safety factor~${F_s}$ is specified as ${F_s}=1.25$. 
    We can then examine that the solutions are in the asymptotic range of convergence 
    \begin{equation}
    \frac{{GC{I_\text{m2c}}}}{{{GCI_\text{f2m}\text{ }}{r_\text{eff}^p}}} = 0.982959 \text{,} 
    \label{eq:C-Range}
    \end{equation}
    which is approximately one and indicates the solutions are within the asymptotic range of convergence.
    Hence, we determine that the prediction has second-order accuracy. 
    The prediction of $St_\text{midium}$ with the medium mesh is within approximately $3.56\%$ of the asymptotic solution~$St_\text{asymptotic}$.
    For these reasons, the medium mesh is selected as the baseline mesh in this work.

    In the NACA 0012 airfoil case, we also generate three meshes with different resolutions. 
    The first layer spacing in the normal direction for all meshes is chosen $y^+ \approx 0.8$, and the growth rate in the boundary layer is 1.15. 
    The meshes are locally refined in the separation and shock motion areas.
    All the generated meshes are simulated at $Ma=0.7, \alpha=5.0^\circ, Re_C=3.0\times {10^6}$ without control. 
    Table~\ref{tab:MeshNACA0012} summarizes the details of mesh information and the shock buffeting frequency with the corresponding mesh. 
    The effect of the grid resolution on the~$St$ of shock buffeting frequency is plotted in Figure~\ref{figA:CylinderMeshStudy} (b), which shows that the~$St$ gradually converges to an asymptotic solution~$St_\text{asymptotic}$ with the mesh density increases.

\begin{table}
    \caption{The mesh information for flows over NACA 0012 airfoil}
    \label{tab:MeshNACA0012}
    \centering
    \begin{tabular}{p{2.5cm}p{2cm}p{2cm}p{2cm}}
    \hline
    \hline
                    & Coarse   & Medium  & Fine     \\
    \hline          
    Number of cells &  43,569  & 85,018  & 170,781  \\
    $St$ $^a$       &  0.05021 & 0.05410 & 0.05619  \\
    \hline
    \hline
    \multicolumn{4}{l}{$^a$ The reference length $L_\text{ref}$ is specified as the chord length of the airfoil ($C$). } 
    \end{tabular}    
\end{table}

    Similar to the GCI analysis for the cylinder flows, the theoretical precision is determined to be second order since the order ($p$) of convergence is $p=1.7813$.
    The predictions of $St$ approach $St_\text{asymptotic}=0.05862 $ as the mesh spacing reduces.
    The mesh convergence index for both the fine mesh solution ($GCI_\text{f2m}$) and medium mesh solution ($GCI_\text{m2c}$) is $ GCI_\text{f2m}=0.053985$ and $ GCI_\text{m2c}=0.104360$, respectively.   
    The asymptotic range of convergence is $ \frac{{GC{I_\text{m2c}}}}{{GC{I_\text{f2m  }}{r_\text{eff}^p}}} = 1.038632 $, which indicates the solutions are within the asymptotic range of convergence.
    Moreover, the prediction~$St_\text{medium}$ with the medium mesh is within approximately $5.3\%$ of the asymptotic solution $St_\text{asymptotic}$.  
    Given these results, the medium mesh is selected as the baseline mesh for the simulations of the NACA 0012 airfoil.

\section{Validation of the CFD solver} \label{sec:AppendixC}

    In this section, we perform the validation process to ensure the accuracy and reliability of the CFD solver. 
    This process involves comparing the CFD prediction with available experimental data for the two investigated cases: flows around a cylinder and the NACA0012 airfoil.

\begin{figure}[!htb]
    \centering    
    \subfloat[$C_p$-distribution]{
    \includegraphics[height=0.36\textwidth]{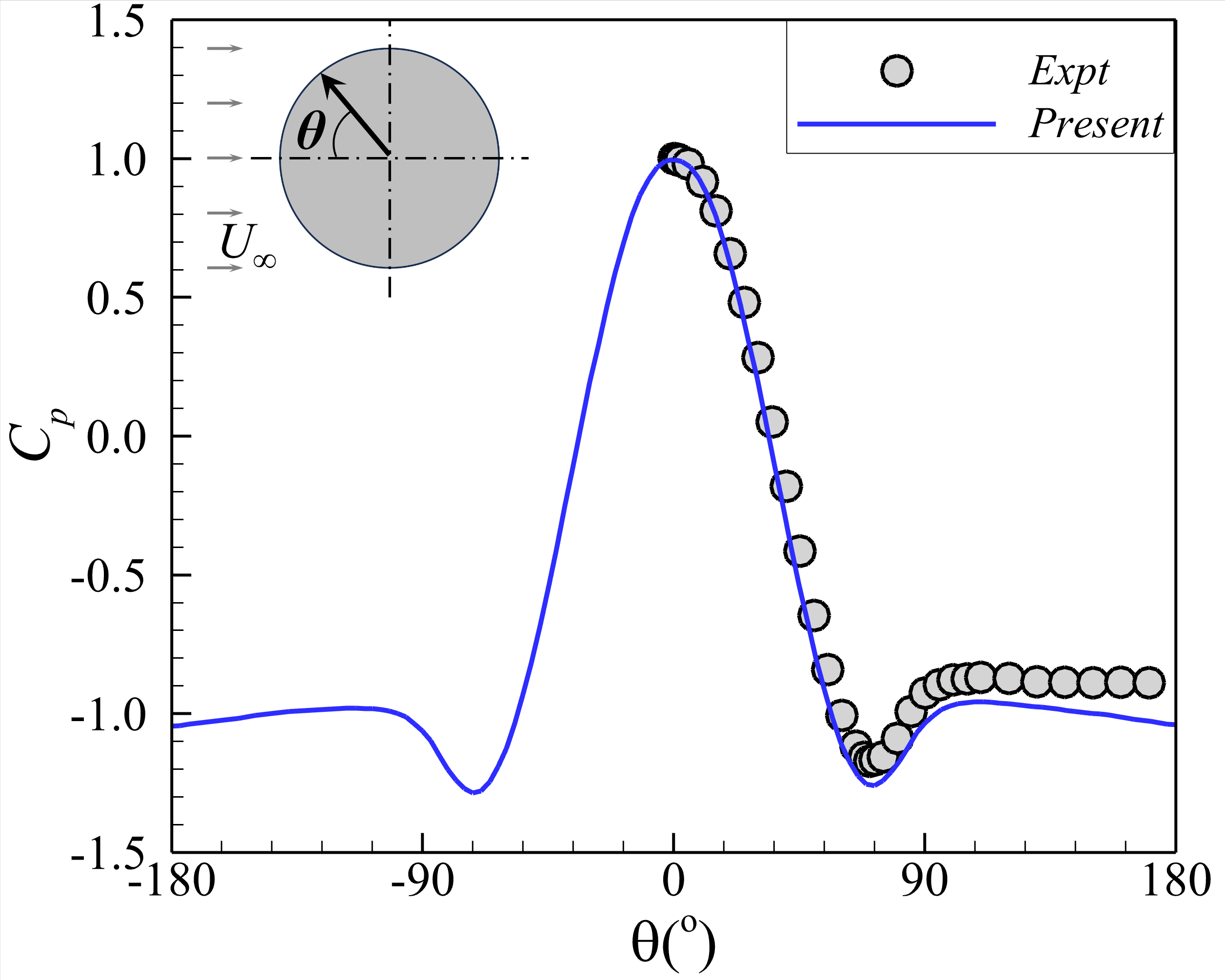} }
    \hspace{1mm}
    \subfloat[$u_{rms}$ at $x/D=1.54$]{
    \includegraphics[height=0.36\textwidth]{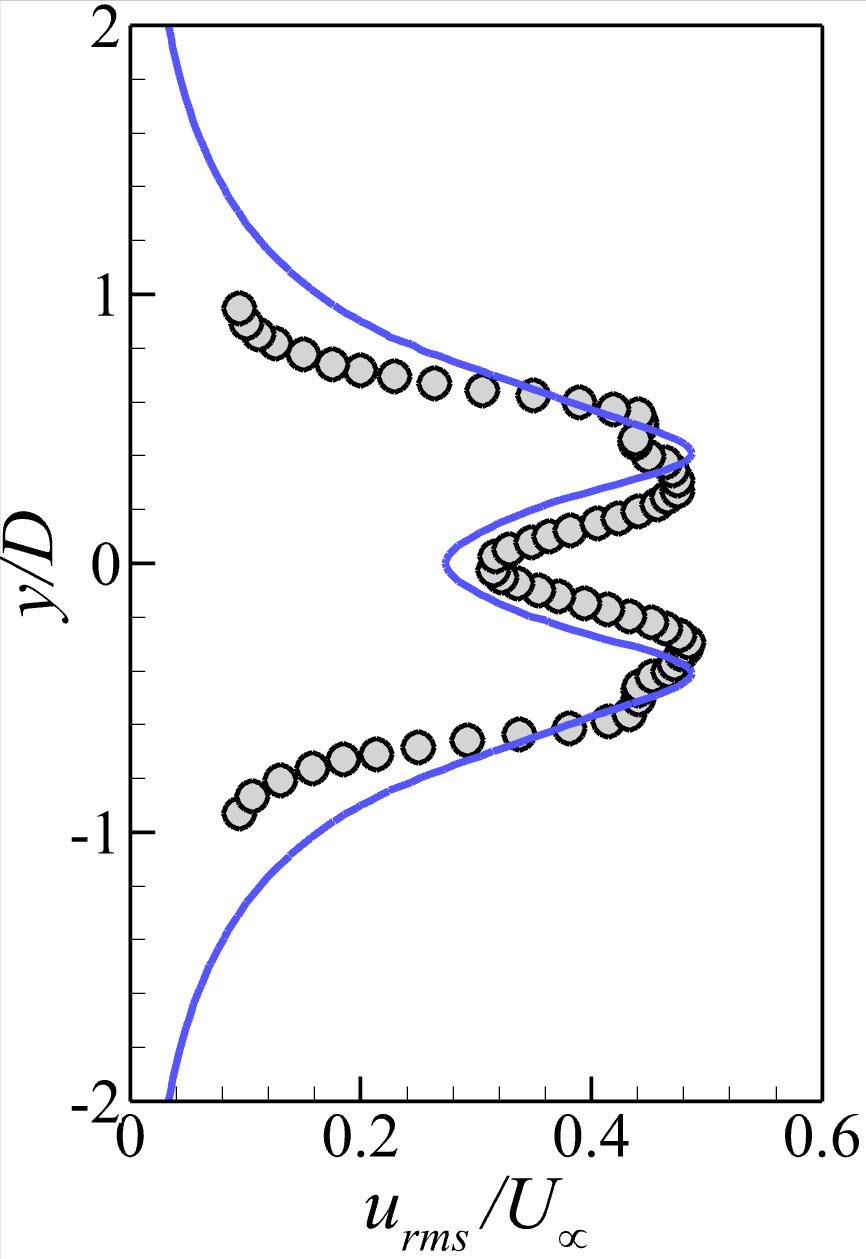} }
    \hspace{1mm}
    \subfloat[$u_{rms}$ at $x/D=2.02$]{
    \includegraphics[height=0.36\textwidth]{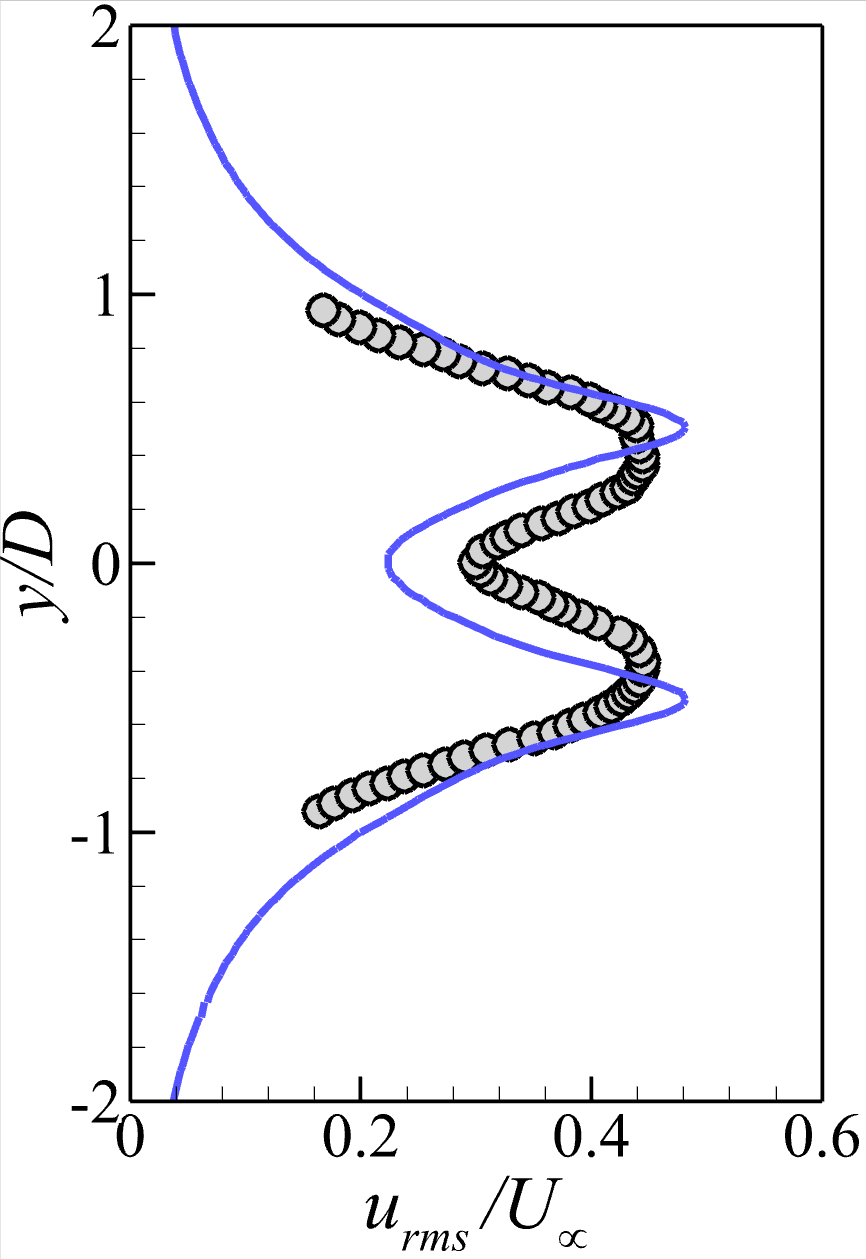} }
    \caption{ Comparison of (a) the averaged $C_p$ distribution, and (b,c) RMS velocity with available experimental data for the circular cylinder case. }
    \label{figC:CylinderExpt}
\end{figure}

    For the flow around a cylinder at Reynolds number $Re_D=3900$, simulations were performed using the medium grid based on the verification in~\ref{sec:AppendixA}. 
    The results are presented in Table~\ref{tab:V&VCylinder} and Figure~\ref{figC:CylinderExpt}, which show the comparison between the CFD prediction and experimental data.
    Table~\ref{tab:V&VCylinder} illustrates flow predictions in the mean drag coefficient ($\langle{C_d}\rangle _\text{mean}$), the mean base pressure coefficient ($C_{P_b}$), the flow separation angle ($\theta_\text{sep}$), and Strouhal number ($St$). 
    The simulation results show good agreement with experimental data~\cite{kravchenko2000numerical,ong1996velocity,norberg2002pressure}.
    Figure~\ref{figC:CylinderExpt} presents the predicted pressure coefficient ($C_p$) distribution, which closely matches experimental data~\cite{norberg2002pressure}.
    Figure~\ref{figC:CylinderExpt} (b) and (c) show the RMS-velocity profiles at $x/D=1.54$ and $x/D=2.02$, which also exhibit good agreement with experiments~\cite{parnaudeau2008experimental}.

\begin{table}[!htb]
    \caption{Flow prediction from cylinder flow computations at $Re_{D}=3900$.  }
    \label{tab:V&VCylinder}
    \centering
    \begin{tabular}{p{2.6cm}p{2cm}p{2cm}p{2cm}p{2cm}}
    \hline
    \hline
                & $\langle{C_d}\rangle _\text{mean}$ & $-C_{P_b}$ & $\theta_\text{sep}$ & $St$ \\
    \hline
    CFD     & $1.088$        & $1.023$        & $88.3^\circ$        & $0.223$  \\
    Experiments~$^a$ & $0.98\pm 0.05$ & $0.9\pm 0.005$ & $84.0^\circ\pm 0.5$ &  $0.215 \pm 0.005$  \\
    \hline
    \hline
    \multicolumn{5}{p{12.5cm}}{ $^a$ $\langle{C_d}\rangle _\text{mean}$ and $St$ are from Refs.~\cite{kravchenko2000numerical} and~\cite{ong1996velocity} at $Re_{D}=3900$, $-C_{P_b}$ and $\theta_\text{sep}$ are from Ref.~\cite{norberg2002pressure} at $Re_{D}=4000$. } 
    \end{tabular}
\end{table}

\begin{figure}[!htb]
    \centering
    \includegraphics[height=0.36\textwidth]{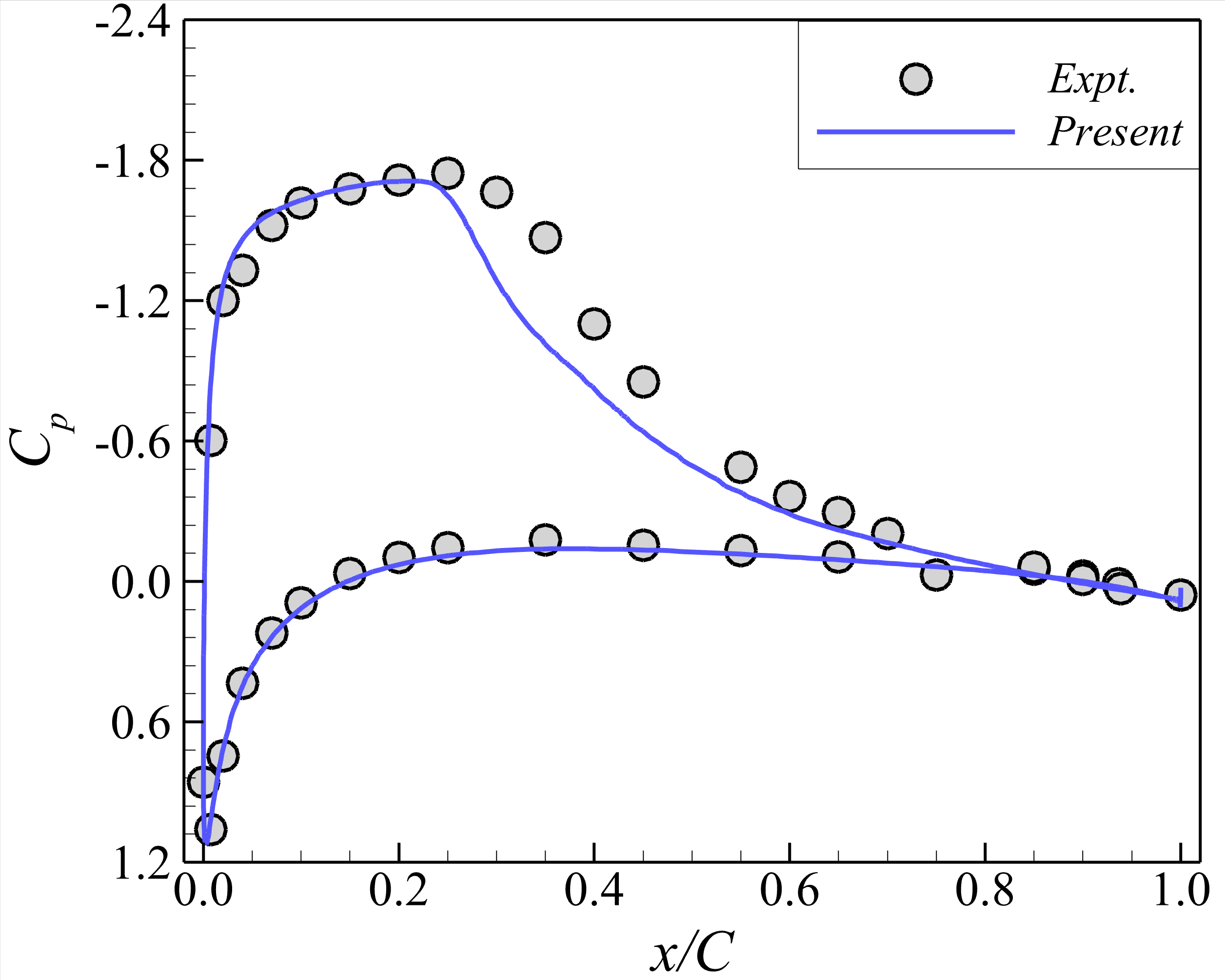} 
    \caption{The averaged $C_p$ distribution compared with experiments~\cite{doerffer2010unsteady} for transonic shock buffeting over NACA 0012 airfoil. }
    \label{figC:V&VNACA0012}
\end{figure}

    The second validation case is the transonic shock buffeting over the NACA0012 airfoil at $Ma=0.7$, $\alpha=5.5^\circ$, and $Re_c=3.0\times 10^6$ using a medium grid as illustrated in~\ref{sec:AppendixA}. 
    The results are displayed in Table~\ref{tabc:V&VNACA0012} and Figure~\ref{figC:V&VNACA0012}, showing the comparison between computational and experimental data. 
    Table~\ref{tab:V&VCylinder} presents the predicted $St=0.05410$, compared to the experimental data of $0.05597$, with a relative error of $3.341\%$.
    Figure~\ref{figC:V&VNACA0012} illustrates the averaged pressure coefficient ($C_p$) distribution, where the predictions show good agreement with the experimental data from Ref.~\cite{doerffer2010unsteady}.

    Conclusively, these comparisons validate the reliability of the CFD solver for predicting key flow characteristics in both circular cylinder and NACA 0012 airfoil scenarios, providing confidence in its application for the flow control optimization in this study.

\begin{table}
    \caption{Prediction of shock buffeting frequency ($St$) for flows over NACA 0012 airfoil.  }
    \label{tabc:V&VNACA0012}
    \centering
    \begin{tabular}{p{1.5cm}p{2cm}p{2cm}p{2cm}}
    \hline
    \hline
               &  CFD   & Experiments~$^a$  & Relative Error \\
    \hline    
    $St$       &  0.05410   &  0.05597     &    3.341 \%     \\
    \hline
    \hline
    \multicolumn{4}{l}{$^a$ The experiments is from Ref.~\cite{doerffer2010unsteady}}
    \end{tabular}    
\end{table} 

\section*{Acknowledgment}
    This work is supported by the NSFC Basic Science Center Program for ``Multiscale Problems in Nonlinear Mechanics'' (No. 11988102), the National Natural Science Foundation of China (Nos. 12102439 and 12102435), and the China Postdoctoral Science Foundation (No. 2021M703290).

\end{document}